# Adversarial swarms as dynamical systems: a sharp outcome transition at population parity and short predictability horizons

Soham Gupta[1a)] and John Baker[1]
[1]Department of Aerospace Engineering and Mechanics, The University of Alabama, Tuscaloosa, Alabama 35487, USA
[a)]Corresponding Author: sgupta5@crimson.ua.edu

## LEAD PARAGRAPH

Swarms of simple agents, whether starlings, fish, or small drones, can act together without a leader. Much less is known about what happens when two swarms with opposite goals meet. Here we study a minimal physics-based encounter: an attacking swarm tries to reach a fixed target while evading a defending swarm that circles the target and intercepts any attacker that comes close. Agents that collide are removed, so the engagement has a definite outcome. By simulating thousands of encounters we find that the result is decided almost entirely by relative numbers: when attackers are even slightly more numerous than defenders the target almost always falls, and when they are slightly fewer it almost always survives. Near parity the engagement lasts longest and is least predictable: once the swarms make contact, a displacement of a single agent by one part in a hundred million grows to the size of the swarm within about one time unit, and in a few percent of encounters it changes the winner. The defending swarm, held near the target, is the more sensitive of the two. These results connect swarm-on-swarm engagements to the physics of interacting self-propelled particles and give a quantitative baseline for the design of defensive swarms.

## ABSTRACT

We study an agent-based model of two adversarial swarms as a dynamical system. Attackers seek a fixed goal while evading defenders; defenders are bound to the goal by a Morse potential and pursue attackers that enter a sphere of influence around it. Agents are self-propelled, interact through Morse and inverse-distance potentials, and are removed on contact, so every encounter ends with a dominant swarm. For 22 population ratios $N_A/N_D$ between 0.05 and 20 (1000 trials each), the probability that the attackers dominate jumps from 0.046 at $N_A/N_D = 1$ to 0.971 at 1.1, and both outcomes occur only for $0.24 \leq N_A/N_D \leq 1.33$. Without attrition, the center-of-mass motion of both swarms passes the 0-1 test for chaos at every ratio, the goal-bound defenders diverging faster and carrying more entropy. With attrition, twin trajectories differing by a $10^{-8}$ displacement separate to the swarm scale within 0.9 to 1.4 time units, and the displacement changes the winner in about 4% of trials near parity; before contact the attackers damp the perturbation and the defenders amplify it. Over the 22,000-trial ensemble the defenders' largest Lyapunov exponent is positive in 97 to 100% of trials across the contested band, whereas the attackers relax to a regular orbit once it resolves. A finite sensing radius leaves the outcomes far from parity unchanged but produces stand-offs near parity and, at the smaller radius, replaces attrition by fast breaches and stand-offs: the sharp transition is a property of long-range sensing.

## 1. INTRODUCTION

Swarms of autonomous agents are found throughout nature, from ant colonies[1] and starling flocks[2] to fish schools[3] and human crowds,[4] and they are increasingly built by engineers from unmanned aerial vehicles,[5] robots,[6] and spacecraft.[7] A swarm gains survival and foraging advantages from collective motion,[8,9] and much of the modeling effort of the past three decades has been devoted to understanding how simple local rules produce such collective behavior. Reynolds[10] showed that attraction, collision avoidance and velocity alignment suffice; Vicsek et al.[11] introduced the self-propelled particle model whose kinetic phase transition has been studied extensively;[12] and D'Orsogna et al.[13,14] showed that self-propelled particles with Rayleigh friction interacting through a Morse potential form rotating mills, flocks and rings in free space, with a sharp distinction between H-stable and catastrophic parameter regimes. Dynamical models of this second-order type, including the Cucker–Smale family,[15-19] have the advantage that forces, inertia and energy are explicit, so the swarm can be analyzed with the tools of nonlinear dynamics.

Swarms rarely exist alone. In nature, groups of predators hunt groups of prey, from chimpanzees and red colobus monkeys[20] and lions and zebras[21] to dolphins, killer whales and herring;[22,23] group hunting is now recognized as a widespread mechanism across vertebrate taxa.[24] In engineering, the defense of a high-value asset against an attacking swarm has become a design problem in its own right, approached through optimal control,[25,26] interception and herding assignments,[27] reinforcement learning,[28] and live experiments with counter-swarms.[29,30] These studies ask how a defending team should act; they generally take the agent-level dynamics as a given control plant rather than as a physical system whose collective behavior is itself the object of study.

The physics of interacting swarms has been developed mainly for symmetric encounters. Zhdankin and Sprott[31] found chaos, quasi-periodicity and singular behavior in a minimal predator–prey swarming model; Angelani[32] and Lin and Abaid[33] studied collective predation with kinematic agents; Kamimura and Ohira[34] and Wang et al.[35] used lattice models of group chase and escape; and Chakraborty et al.[36,37] showed that the survival of a prey swarm depends non-monotonically on the range of cooperative interactions and on the predator-to-prey mass ratio. For two swarms of self-propelled particles, Kolon and Schwartz[38] described mutual capture and milling after collision; Hindes et al.[39] derived the critical coupling below which two colliding swarms merely scatter rather than forming a combined milling state; Kamimoto, Hindes and Schwartz[40] showed that the post-collision milling state becomes chaotic as the repulsive force strength and length scale are varied, using the 0-1 test for chaos; and Hindes et al.[41] extended the framework to countermoving swarms that are redirected by the collision. On the single-swarm side, Medeiros and Feudel[42] showed that transitions between collective states of a Morse–Rayleigh swarm pass through chaotic dynamics associated with fractal basin boundaries, and Taylor and Nowzari[43] examined how the removal of agents by catastrophic collisions alters the emergent behavior. The defender–intruder problem has been treated for one defender and one intruder, where motion patterns and a phase transition between them have been characterized.[44,45]

What is missing between these two bodies of work is a dynamical-systems account of an asymmetric engagement: two swarms with different objectives, one of them bound to a fixed goal, in which agents are removed on contact so that every encounter has an outcome. The questions such a model raises are different from those of symmetric collisions. How does the outcome depend on the relative size of the swarms, and how sharp is that dependence? Is the engagement chaotic, and if so, which swarm carries the chaos? How does the complexity of the collective motion vary with the population ratio? And does the answer depend on the assumption, common to Morse-type models, that every agent senses every other?

In this paper we address these questions with a model in which an attacker swarm seeks a goal in unbounded two-dimensional space while a defender swarm, held near the goal by a Morse well, pursues attackers that enter a circular sphere of influence. Intra-swarm forces are Morse forces in the H-stable regime;[13] inter-swarm attraction and repulsion derive from inverse-distance potentials;[46] every agent is self-propelled with Rayleigh friction;[47] and agents that approach within a small contact distance are removed. The model is solved with a fourth-order Runge–Kutta scheme whose time step is fixed by a convergence study, for 22 population ratios between 0.05 and 20 and 1000 random initial conditions per ratio. The contributions are:

1. An outcome phase diagram with confidence intervals. The probability of attacker dominance rises from below 5% to above 97% between $N_A/N_D = 1.0$ and 1.1, and both outcomes are possible only in the band $0.24 \leq N_A/N_D \leq 1.33$, where the engagement also lasts longest.
2. A quantitative characterization of chaos and predictability. In sustained interaction both swarms are chaotic by the 0-1 test, with the goal-bound defenders diverging faster and carrying more entropy for $N_A/N_D \geq 0.5$. In attritional engagements, twin trajectories show that the attackers damp a small perturbation before contact while the defenders amplify it, that the engagement becomes unpredictable within about one time unit of its start, and that a $10^{-8}$ displacement of one agent flips the outcome in a few percent of trials near parity. The ensemble of 22,000 Wolf exponents, which found the defenders chaotic far more often than the attackers, is explained by the relaxation of the attackers to a regular orbit once the engagement resolves.
3. A complexity measure. Multiscale entropy, defined fully here, shows the defenders to be the more complex swarm at every population ratio in the original ensemble and for $N_A/N_D \geq 0.5$ in the sustained-interaction variant.
4. A sensitivity study with a finite sensing radius, which replaces the all-to-all coupling of the original model. The outcomes far from parity are unchanged, but near parity a sensing radius of 0.5, about the radius of the sphere of influence, produces stand-offs in about 40% of the trials, and a radius of 0.25 changes the character of the engagement from attrition to breach-or-stand-off: the sharp transition is a property of long-range sensing. We also identify, and exclude, trials in which the initializer placed two agents at the same point.

Section 2 defines the model and the numerical method, Section 3 the analysis tools, Section 4 presents the results, Section 5 discusses them in relation to the interacting-swarm literature and to swarm-defense design, and Section 6 provides a concise listing of the conclusions of the study. A derivation of the Cartesian force components is given in the Appendix.

## 2. MODEL AND NUMERICAL METHOD

### 2.1 Equations of motion

Two swarms interact in unbounded two-dimensional Euclidean space: $N_A$ attackers, whose objective is to reach a fixed point called the goal, and $N_D$ defenders, whose objective is to keep them from it. Each agent is a point mass with position X and velocity V. The equations of motion, obtained from Newton's second law, are

$$\ddot{\vec{X}}_{A_i} = \dot{\vec{V}}_{A_i} = \frac{1}{m_{A_i}} \Big( \sum_{\substack{A_i=1 \\ A_i \neq A_k}}^{N_A} -(\vec{\nabla}\varphi_{A,ik}) + \sum_{A_i=1}^{N_D} -\vec{\nabla}(k_{rep} r_{ij}^{-1}) \qquad (1)$$

$$-\vec{\nabla}(-k_{obj} r_{iG}^2) + (\alpha_A - \beta_A |\vec{V}_{A_i}|^2) \vec{V}_{A_i}$$

$$\dot{\vec{X}}_{A_i} = \vec{V}_{A_i} \qquad (2)$$

$$\ddot{\vec{X}}_{Dj} = \dot{\vec{V}}_{Dj} = \frac{1}{m_{Dj}} \Big( \sum_{\substack{Dj=1 \\ Dj \neq Dh}}^{N_D} -(\vec{\nabla}\varphi_{D,jh}) + \sum_{Dj=1}^{N_A} -\vec{\nabla}(-k_{att} r_{ji}^{-1}) - \vec{\nabla}(\varphi_{jG}) + (\alpha_D - \beta_D |\vec{V}_{Dj}|^2) \vec{V}_{Dj} \qquad (3)$$

$$\dot{\vec{X}}_{Dj} = \vec{V}_{Di} \qquad (4)$$

subject to the initial positions and velocities of every agent. The first term of Eqs. (1) and (3) is the intra-swarm force, the negative gradient of the scaled Morse potential[13,48]

$$\varphi_{ij} = C \exp\left(-\frac{|r_{ik}|}{l}\right) - \exp\left(-\frac{|r_{ik}|}{l}\right) \qquad (5)$$

where C sets the depth of the repulsive well and l the ratio of repulsive to attractive length scales. With C > 1 and l > 1 the repulsion acts at short range and the attraction at long range, which keeps a swarm cohesive without collapse; we choose C and l for both swarms in the H-stable regime $Cl^2 > 1$ of D'Orsogna et al.,[13,14] in which the swarm remains bounded as the number of agents increases. Because the Morse force decays exponentially, it imposes an effective sensing range on each agent without an explicit cutoff; Section 4.5 tests this assumption directly.

The second term is the inter-swarm force. Both the repulsion felt by an attacker from the defenders and the attraction felt by a defender toward the attackers derive from the inverse-distance potential of Espitia and Sofrony,[46]

$$\varphi_A = \pm k_{rep/att} r_{ij}^{-1} \qquad (6)$$

with a positive constant $k_{rep}$ for repulsion and a negative constant $k_{att}$ for attraction. The defenders are attracted to an attacker only while that attacker is inside the sphere of influence, a circle of radius $R_{th}$ around the goal:

$$\vec{F}_{DA,A} = -\nabla(-k_{att} r_{ij}^{-1}) \text{ if } r_{ij} \le R_{th} \qquad (7)$$

$$\vec{F}_{DA,A} = 0 \text{ if } r_{ij} > R_{th} \qquad (8)$$

This gating keeps the defenders near the goal and, as we show below, is the source of the switching that makes their motion chaotic. The inverse-distance forces are capped at a local threshold distance (Table 1), below which they are held at the threshold value; this prevents the forces from diverging during close engagements and the same threshold is used for both swarms.

The third term of Eq. (1) is a linear goal force on each attacker, derived from a quadratic potential[46] and increasing with the distance to the goal. The third term of Eq. (3) binds each defender to the goal through a second scaled Morse potential

$$\varphi_{jG} = C_{DG} \exp\left(-\frac{|r_{jG}|}{l_{DG}}\right) - \exp(-|r_{jG}|) \qquad (9)$$

with its own constants $C_{DG}$ and $l_{DG}$, again in the H-stable regime, so that a defender is repelled from the goal at short range and attracted at long range and therefore orbits it. The last term of Eqs. (1) and (3) is the Rayleigh self-propulsion and friction force,[13,47]

$$\vec{F}_{A,SF,i} = (\alpha_A - \beta_A |\vec{V}_{A,i}|^2) \vec{V}_{A,i} \qquad (10)$$

$$\vec{F}_{D,SF,i} = (\alpha_D - \beta_D |\vec{V}_{D,i}|^2) \vec{V}_{D,i} \qquad (11)$$

which drives every agent toward the speed $(\alpha/\beta)^{1/2}$ in the absence of other forces. Agents have unit mass.

### 2.2 Contact, termination and outcome

Two rules make the encounter terminate. If an attacker and a defender come within the contact distance $d_c = 10^{-4}$ of each other, both are removed from the simulation for the rest of the run; the number of agents removed in a single event is not necessarily two, because several agents can be within $d_c$ at the same step. If an attacker comes within $d_c$ of the goal, the goal is breached. A run ends, and a dominant swarm is declared, at the first of the following events: the goal is breached or the last defender is removed (attackers dominant); the last attacker is removed, or the maximum simulation time $T_{max} = 100$ is reached with the goal intact (defenders dominant). If both swarms are annihilated at the same step the defenders are dominant, because the goal was protected. These rules make the model a hybrid dynamical system: smooth second-order dynamics between discrete events at which the state dimension drops.

### 2.3 Initial conditions, parameters and integration

The goal is fixed at (0.5, −0.5), the center of the fourth quadrant of the square $[-1, 1]^2$ (Fig. 1). The attackers are initialized in the second quadrant with a scrambled Sobol sequence,[49] which spreads the initial positions evenly, and the defenders inside the sphere of influence around the goal, $R_{th}$ = 0.508, at a radius and an angle drawn uniformly at random; initial velocity components of both swarms are drawn uniformly from [−0.1, 0.1]. All other parameters are listed in Table 1. They were chosen by numerical experiment so that each swarm, in isolation, forms a stable cohesive state and so that the engagement produces a balance between evasion and interception; the inter-swarm constants, the contact distance and the force thresholds are identical for both swarms so that neither is favored by the force law itself.

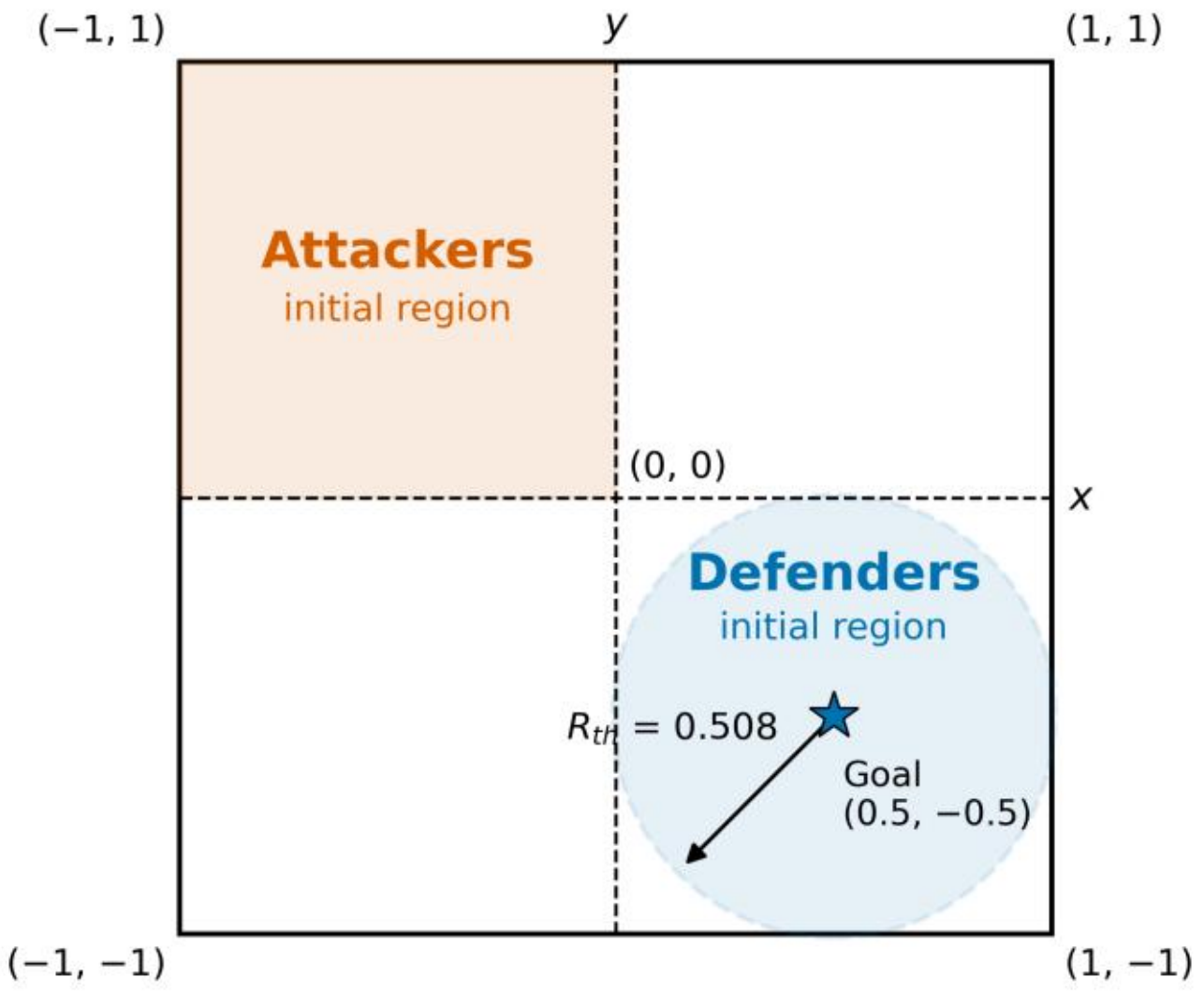

**FIG. 1.** Simulation domain. Attackers are initialized in the second quadrant and defenders inside the sphere of influence (dashed circle, radius $R_{th}$ = 0.508) around the goal at (0.5, −0.5).

**TABLE 1.** Model parameters (suffix A, attackers; D, defenders).

| Constants | Values | Description |
|---|---|---|
| $C_A$ | 1.32 | Depth of the potential well for pairwise scaled Morse potential for Attackers. |
| $l_A$ | 1.50 | The ratio of the repulsive to attractive length scales of Attacker agents. |
| $C_D$ | 1.38 | Depth of the potential well for pairwise scaled Morse potential for Defenders. |
| $l_D$ | 2.00 | The ratio of the repulsive to attractive length scales of Defender agents |
| $k_{rep}$ | 0.054 | Constant for pairwise inter-swarm repulsive potential between Attackers and Defenders. |
| $k_{att}$ | 2.82 | Constant for pairwise inter-swarm attraction potential between Defenders and Attackers. |
| $k_{obj}$ | 10 | Constant for pairwise inter-swarm Goal potential between Attackers and the Goal. |
| $C_{DG}$ | 10.03 | The ratio of attraction and repulsive potential strength for scaled Morse potential between a Defender agent and the Goal. |
| $l_{DG}$ | 0.5 | The ratio of the length scale of attraction and repulsion pair for scaled Morse potential between a Defender agent and the Goal. |
| $\alpha_A$ | 1 | Constant of the self-propulsion force for Attackers. |
| $\beta_A$ | 1 | Coefficient of the Rayleigh friction force for Attackers. |
| $\alpha_D$ | 1 | Coefficient of the self-propulsion force for Defenders. |

| Constants | Values | Description |
|---|---|---|
| $\beta_D$ | 1 | Coefficient of the Rayleigh friction force for Defenders. |
| $R_{th}$ | 0.508 | Influence radius for Defenders. |
| $m_A$ | 1 | Mass of Attacker agents. |
| $m_D$ | 1 | Mass of Defender agents |
| Goal breach criteria | 1E-4 | Threshold distance between Attacker agents and the Goal for considering the Goal to be breached. |
| Agent compromise criteria | 1E-4 | Threshold distance between individual Attacker agents and Defender agents to be considered collided (hence dead). |
| Repulsive Force Local Cut-off for Attacker | 0.05 | The minimum distance beyond which the repulsive force between an Attacker and a Defender agent is treated as constant. |
| Attraction Force Local Cut-off for Defenders | 0.05 | The minimum distance beyond which the attractive force between a Defender and an Attacker agent is treated as constant. |

Equations (1) to (4) are integrated with a fourth-order Runge–Kutta scheme written in C++.[50] The integrator was verified against a mass–spring–damper system and its order of accuracy confirmed numerically. The time step was fixed for each population ratio by a convergence study in which the step was halved until the position vectors of every agent changed by less than 1% over the run; the resulting steps are listed in Table 2 and range from $5 \times 10^{-5}$ at $N_A/N_D = 0.05$ to $2 \times 10^{-6}$ at $N_A/N_D = 20$, the smallest steps being required when many attackers crowd the goal.

### 2.4 Monte Carlo design

Twenty-two population ratios were simulated (Table 2), from $N_A = 5$, $N_D = 100$ to $N_A = 100$, $N_D = 5$, with the total population held near 105. Each ratio was run 1000 times from independent random initial conditions on the University of Alabama high-performance computing cluster and a desktop workstation. The cumulative mean of the run time changed by less than 0.2% over the last runs of every case, which we took as the convergence criterion for the ensemble. The random initial conditions are the only source of randomness; the dynamics are deterministic.

**TABLE 2.** Case matrix: population sizes, ratio, number of trials, time step, and $T_{max}$.

| Case | NA | ND | NA/ND | Runs | Timestep | Max Simulation Time |
|---|---|---|---|---|---|---|
| 1 | 5 | 100 | 0.05 | 1000 | 5.00E-05 | 100 |
| 2 | 7 | 97 | 0.07 | 1000 | 2.50E-05 | 100 |
| 3 | 10 | 95 | 0.10 | 1000 | 2.50E-05 | 100 |
| 4 | 15 | 90 | 0.16 | 1000 | 2.50E-05 | 100 |
| 5 | 20 | 85 | 0.23 | 1000 | 2.50E-05 | 100 |
| 6 | 25 | 80 | 0.31 | 1000 | 2.50E-05 | 100 |
| 7 | 30 | 75 | 0.4 | 1000 | 2.50E-05 | 100 |
| 8 | 35 | 70 | 0.5 | 1000 | 2.50E-05 | 100 |
| 9 | 40 | 65 | 0.61 | 1000 | 2.50E-05 | 100 |
| 10 | 45 | 60 | 0.75 | 1000 | 2.50E-05 | 100 |
| 11 | 50 | 55 | 0.90 | 1000 | 2.50E-05 | 100 |
| 12 | 50 | 50 | 1 | 1000 | 2.50E-05 | 100 |
| 13 | 55 | 50 | 1.1 | 1000 | 2.50E-05 | 100 |
| 14 | 60 | 45 | 1.33 | 1000 | 2.50E-05 | 100 |
| 15 | 65 | 40 | 1.62 | 1000 | 2.50E-05 | 100 |
| 16 | 70 | 35 | 2 | 1000 | 2.50E-05 | 100 |
| 17 | 75 | 30 | 2.5 | 1000 | 1.00E-05 | 100 |

| Case | NA | ND | NA/ND | Runs | Timestep | Max Simulation Time |
|---|---|---|---|---|---|---|
| 18 | 80 | 25 | 3.2 | 1000 | 5.00E-06 | 100 |
| 19 | 85 | 20 | 4.25 | 1000 | 5.00E-06 | 100 |
| 20 | 90 | 15 | 6 | 1000 | 5.00E-06 | 100 |
| 21 | 95 | 10 | 9.5 | 1000 | 5.00E-06 | 100 |
| 22 | 100 | 5 | 20 | 1000 | 2.00E-06 | 100 |

# 3. ANALYSIS METHODS

### 3.1 Center-of-mass series and phase-space reconstruction

The collective motion of each swarm is characterized by the position of its center of mass, $X_{cm}(t) = N^{-1} \Sigma X_i(t)$, averaged over the agents still active at time t; the velocity of the center of mass is defined in the same way. The removal of agents produces jumps in $X_{cm}$, which are part of the dynamics of the engagement and are retained. The series are sampled every $10^{-3}$ time units. The x component of $X_{cm}$ is embedded with the method of delays: the delay τ is the first minimum of the time-delayed mutual information[51] and the dimension m the smallest for which the fraction of false nearest neighbors[52] falls below 1%, using the implementation of Chen and Yang.[53]

### 3.2 Largest Lyapunov exponent over the ensemble

For every trial of every case the largest Lyapunov exponent $\lambda_1$ of the embedded center-of-mass series was computed with the algorithm of Wolf et al.,[54]

$$\Lambda_1 = \frac{1}{M_{LE} t_{evol}} \sum_{k=0}^{M} \ln \frac{L_{evol}^{(i)}}{L_o^{(i)}} \tag{12}$$

in which the separation $L_0$ between a point and its nearest neighbor evolves to $L_{evol}$ after a time $t_{evol}$ and M replacement steps are followed. These 22,000 estimates are the only chaos measure that covers the whole case matrix. They have a limitation that should be stated: an engagement that ends after 7 to 15 time units yields a series of only 10 to 30 characteristic periods, during which agents are removed and the center of mass jumps, so $\lambda_1$ is a finite-series estimate of the sensitivity to initial conditions during the engagement rather than an asymptotic exponent. Two measures that do not share this limitation were therefore added.

### 3.3 Twin trajectories: predictability horizon and outcome sensitivity

For the baseline engagements of the sensitivity study (Section 3.7) every trial was rerun from the same initial condition with the x coordinate of one attacker and one defender displaced by $\delta = 10^{-8}$. The distance d(t) between the centers of mass of the original and the displaced run was recorded for each swarm. Three quantities were extracted: the growth rate of ln d(t) before the first removal of an agent, which measures the sensitivity of the pre-contact approach; the predictability horizon $t_h$, the first time at which d exceeds $10^{-3}$, that is, the time at which the two realizations of the same engagement have become macroscopically different; and whether the twin runs ended with the same dominant swarm. These quantities require no embedding and are defined for transient dynamics. The attritional dynamics are not smooth (agents are removed and the inter-swarm forces are capped at short range), so d(t) after contact grows in steps rather than exponentially, and we report horizons rather than exponents.

### 3.4 The 0-1 test and recurrence plots for non-attritional engagements

The asymptotic character of the interacting dynamics was examined in a variant of the model in which agent removal and termination are switched off, so that the two swarms interact continuously for 150 time units at the same population ratios; the first 30 time units were discarded as transient and the remaining series, sampled every 0.01 time units, were embedded as in Section 3.1. On these stationary series $\lambda_1$ was estimated with the method of Rosenstein et al.[55] from the mean logarithmic divergence of nearest neighbors, the slope being fitted between 10% and 60% of the rise of the divergence curve toward saturation. Before it was applied to the swarm series, the implementation was validated on two reference signals whose exponents are known: the x component of the Lorenz system at $\sigma = 10$, $\rho = 28$, $\beta = 8/3$, the standard benchmark for a chaotic time series, with $\lambda_1 = 0.906$,[56] and a quasi-periodic sum of two incommensurate sinusoids, for which $\lambda_1 = 0$. The implementation gives $\lambda_1 = 0.88$ to 1.11 for the Lorenz series sampled at 0.01 with 4000 to 30,000 points and 0.03 for the quasi-periodic signal, so it recovers a known positive exponent to within the scatter expected for series of this length and returns a value near zero for regular dynamics. Chaos was then tested with the 0-1 test of Gottwald and Melbourne,[57,58] which does not require phase-space reconstruction. For a scalar observable $\varphi(j)$, $j = 1, \ldots, N$, and a constant $c \in (\pi/5, 4\pi/5)$, the translation variables

$$p_c(n) = \sum_{j=1}^{n} \phi(j)\cos(jc), q_c(n) = \sum_{j=1}^{n} \phi(j)\sin(jc) \tag{13}$$

behave diffusively for chaotic dynamics and remain bounded for regular dynamics. The modified mean square displacement $D_c(n) = M_c(n) - V_{osc}(c, n)$, with $M_c(n)$ the mean of $[p_c(j+n) - p_c(j)]^2 + [q_c(j+n) - q_c(j)]^2$ over j and $V_{osc}$ the oscillatory term of Ref. 58, grows linearly in n for chaos, and $K_c$ is the correlation coefficient between n and $D_c(n)$ for $n \leq N/10$. The test statistic K is the median of $K_c$ over 100 values of c; $K \approx 1$ indicates chaos and $K \approx 0$ regular dynamics. The series was sub-sampled at $\tau/2$ before the test to avoid oversampling.[58] The implementation was validated in the same way before use: it gives $K = 0.00$ for a sinusoid and $K = -0.02$ for the quasi-periodic signal, both regular, and $K = 1.00$ for the Lorenz x component (25,000 points) and $K = 1.00$ for white noise, the values expected for chaotic and for noise-like signals. Kamimoto et al.[40] used the same test for colliding swarms. Recurrence plots[59,60] of the same series were used as a qualitative check. With the embedded points $x_i$, the recurrence matrix is

$$R_{s,t} = \Theta(\varepsilon - \| x_s - x_t \|); s,t = 1,2,3,....,N \tag{14}$$

with the threshold $\varepsilon$ fixed at 10% of the diameter of the reconstructed phase space, $\Theta$ the Heaviside function and $\|\cdot\|$ the Euclidean norm. Periodic dynamics produce equally spaced diagonal lines, quasi-periodic dynamics unequally spaced lines, and chaotic dynamics short broken diagonals and isolated points.

### 3.5 Multiscale entropy

Two multiscale entropies are used, one for each ensemble. For the original ensemble (Fig. 11) the complexity of the two-component center-of-mass series (x, y) of each swarm was quantified with the multivariate multiscale sample entropy of Ahmed and Mandic,[61] which extends the multiscale entropy of Costa et al.[62] to multichannel data,

$$M_{SE}(M, \tau, r, N) = -\ln[\frac{B^{m+1}(r)}{B^m(r)}] \tag{15}$$

where M is the vector of embedding dimensions, $\tau$ the vector of time lags, r the tolerance, N the number of samples, and the two frequencies count the composite delay vectors of total length $\Sigma M$ and $\Sigma M + 1$ that match within r. The center-of-mass series of each trial was subsampled to every 20 to 100 integration steps (of order $10^{-3}$ time units; the factor was chosen per case to keep the series length tractable), each component was normalized to unit range and standardized, the series was coarse-grained by non-overlapping averaging at scales 1 to 20, the tolerance was $r = 0.15$ times the sum of the component standard deviations, and the embedding dimension was 2 with unit lag for each component; the implementation was the MATLAB code published with Ref. 61. The ratio of the entropy at scale 20 to that at scale 1 summarizes the growth of entropy with scale for each case: a series with no structure beyond its scale-1 variance has a ratio near 1, and a series with structure across scales, which we take as the signature of complex dynamics, has a ratio well above 1.[62] Because the subsampling factor differs between cases, the ratio is compared across cases but the entropy values themselves are not. The entropy was computed for every trial, and the mean and standard deviation over the trials of each case are reported (1000 trials per case, except 600 at $N_A/N_D = 1.63$ and 905 at $N_A/N_D = 2$; the per-trial values are in the data deposit).

For the non-attritional series of Section 3.4, which are analyzed one coordinate at a time, the modified multiscale entropy of Wu et al.[63] was used, a variant suited to series of moderate length: at scale s the series is coarse-grained by a moving average of window s, and the sample entropy[64] of the coarse-grained series is computed with template length $m = 2$, template spacing s, and tolerance $r = 0.15$ times the standard deviation of the original series,

$$\text{SampEn}(m, r, s) = -\ln\left[\frac{A(m+1,r,s)}{B(m,r,s)}\right] \tag{16}$$

where B and A are the numbers of pairs of templates of length m and $m + 1$ that match within r in the maximum norm. The full curves over scales 1 to 20 (sampling 0.02 time units) and their mean over scales are reported. The two measures are not compared with each other; each is used only within its own ensemble.

### 3.6 Outcome statistics and invalid trials

The probability of attacker dominance at each ratio is estimated from the 1000 trials with a Wilson 95% confidence interval.[65,66] The time to outcome is reported separately for attacker-dominant and defender-dominant trials.

In preparing the new simulations of Sections 3.3, 3.4 and 3.7 we found that for some seeds the scrambled Sobol routine returns the same point twice, so that two attackers start at identical positions; the intra-swarm force then divides by zero and every position becomes non-finite at the first step. Such a run proceeds to $T_{max}$ with no interaction and is logged as a defender win by protection of the goal. These trials are identified by non-finite center-of-mass positions and excluded from all statistics, and replacement seeds are run so that each condition keeps 50 valid trials; they amounted to 6 to 10% of seeds. Whether the original ensemble contained such trials cannot be settled from the surviving records. The per-trial exponents that survive for $N_A = N_D = 50$ are finite in all 1000 trials, which argues that it did not, since a non-finite trajectory gives a non-finite exponent; and the 29 time-limit defender wins recorded at $N_A/N_D = 1.1$ are consistent with the rate of genuine stand-offs in the new valid runs at that ratio (2 of 50, Section 4.5), in which a single surviving defender holds the goal. The identical count of 29 at $N_A/N_D = 1.33$ we cannot check. The original counts are therefore reported as recorded in Section 4.1.

### 3.7 Finite sensing radius

To test the all-to-all coupling implied by the Morse and inverse-distance potentials, the model was rerun with a finite sensing radius $R_s$: an agent experiences the intra-swarm and inter-swarm forces only from agents within $R_s$, and no force from those beyond it. $R_s = 0.5$, which is about the radius of the sphere of influence and half the half-width of the square domain $[-1, 1]^2$, and $R_s = 0.25$ were compared with the original model ($R_s = \infty$) for $N_A/N_D = 0.5$, 0.91, 1.0, 1.1 and 2.0, with 50 trials each, the same time step and the same initial-condition generator. These runs were made on a desktop workstation with a command-line build of the original C++ code, that is, the same integrator compiled without its graphical display so that batches of trials could be run unattended; the only modification to the model adds a single distance test to each pairwise force, and the modified code reproduces the original trajectories exactly when $R_s = \infty$.

### 3.8 Software

The simulation code is the authors' C++ implementation described in Section 2.3. The analysis scripts used in this revision for the embedding, the Rosenstein exponent, the 0-1 test, the twin-trajectory analysis and the multiscale entropy were written in Python 3.12 with NumPy and SciPy with the assistance of a large language model (Claude, Anthropic, 2026), used to accelerate implementation; every routine was verified by the authors against reference signals (white noise, a sinusoid, a quasi-periodic signal and the Lorenz system, Section 3.4) before use. The same tool was used to draft and edit the revised text of this article from the authors' earlier manuscript, data and analyses. The authors reviewed all code and text and take full responsibility for the content.

## 4. RESULTS

### 4.1 Outcome phase diagram

Figure 2(a) shows the probability that the attackers dominate as a function of $N_A/N_D$. For $N_A/N_D \leq 0.17$ the defenders dominate in all 1000 trials; between 0.24 and 0.91 the attackers succeed in 0.1 to 2.0% of trials; at parity they succeed in 4.6% (95% CI 3.5 to 6.1%); at $N_A/N_D = 1.1$ they succeed in 97.1% (95.9 to 98.0%); and for $N_A/N_D \geq 1.63$ they succeed in every trial. The transition is therefore concentrated in a 10% change of the ratio around parity, and both outcomes are observed only in the contested band $0.24 \leq N_A/N_D \leq 1.33$, shaded in the figure. The transition is not a numerical artifact of the case matrix: the two cases that straddle it, $N_A = 50$, $N_D = 50$ and $N_A = 55$, $N_D = 50$, differ by five attackers.

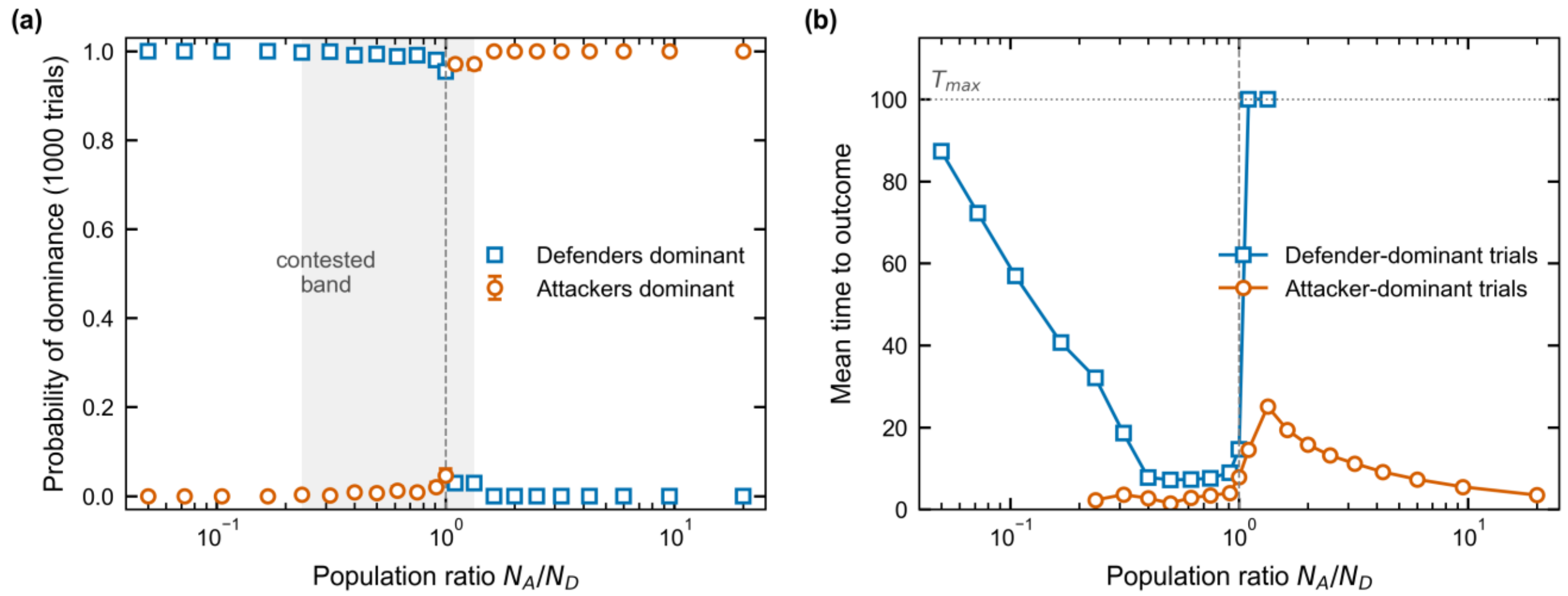

**FIG. 2.** (a) Probability that the attackers dominate (circles) and that the defenders dominate (squares) in 1000 trials, with Wilson 95% confidence intervals; the shaded band marks the ratios at which both outcomes occur. (b) Mean time to outcome for defender-dominant and attacker-dominant trials; the dotted line is $T_{max}$.

The two swarms dominate by different mechanisms. On the defender side of the transition, the attackers succeed by breaching the goal rather than by annihilating the defenders: at parity, 30 of the 46 attacker wins were breaches. On the attacker side the roles reverse: at $N_A/N_D = 2$ and 20 the attackers win every trial, but only 21% and 17% of those wins are breaches, the rest being annihilation of the defender swarm. Defender wins at $N_A/N_D \leq 0.17$ are almost all by protection of the goal until $T_{max}$, which is why the mean time to outcome in Fig. 2(b) is close to $T_{max}$ at the smallest ratios; between 0.4 and 0.9 the defenders annihilate the attackers within 7 to 9 time units. The 29 defender wins recorded at $N_A/N_D = 1.1$ and 1.33 are all time-limit protections. The new baseline ensemble of Section 4.5, from which invalid trials are excluded, gives attacker dominance in 48 of 50 valid trials at $N_A/N_D = 1.1$ (the two exceptions are stand-offs in which a single surviving defender held the goal against six attackers until $T_{max}$) and in 49 of 49 at $N_A/N_D = 2$, against 1 of 50 at parity and 0 of 50 at 0.91 and 0.5, consistent with the original ensemble. The attacker-dominant trials are fastest, 1.5 to 4 time units, on the defender side of the transition, where the rare attacker successes are quick breaches by an attacker that slips past the defenders; the mean time rises to 7.9 at parity and peaks at 25 at $N_A/N_D = 1.33$, where the attackers must wear down a defender swarm of comparable size, and then falls monotonically as the attackers' numerical advantage grows. The engagement is thus longest, and its outcome least predictable, in the band immediately above parity.

### 4.2 Dynamics of representative engagements

The dynamics on either side of the transition are illustrated by single engagements at three ratios (Figs. 3 to 7). At $N_A/N_D = 0.05$ (Fig. 3) the five attackers are met by the defender swarm as they approach the sphere of influence; three are removed in the first few time units, and the two survivors are held outside the sphere by the combined repulsion of the 97 remaining defenders, so that they orbit the goal at a radius of about 0.6 for the rest of the run. The center-of-mass series (Fig. 4) show an irregular transient during the engagement, followed by a large quasi-periodic oscillation of the attacker center with a period of about five time units, while the defender center, after a transient of large excursions as defenders are drawn toward the intruders and return, settles within ten time units to small oscillations about the goal. This is the behavior described as transient chaos by Grebogi, Ott and Yorke,[67] in which trajectories leave a chaotic region after a finite time. The recurrence plots (Fig. 5) show it directly: the attacker plot has isolated points and short diagonal segments during the engagement and blocks of long parallel diagonals, the signature of quasi-periodic motion, once the survivors have settled into their orbit; the defender plot is dense, because the defender center barely moves after the transient, and striped with the diagonals of the residual oscillation. The residence density of the attacker center (Fig. 4(c)) is spread over an annulus outside the sphere of influence; that of the defender center (Fig. 4(f)) is concentrated at the goal, with a tail toward the point of first contact.

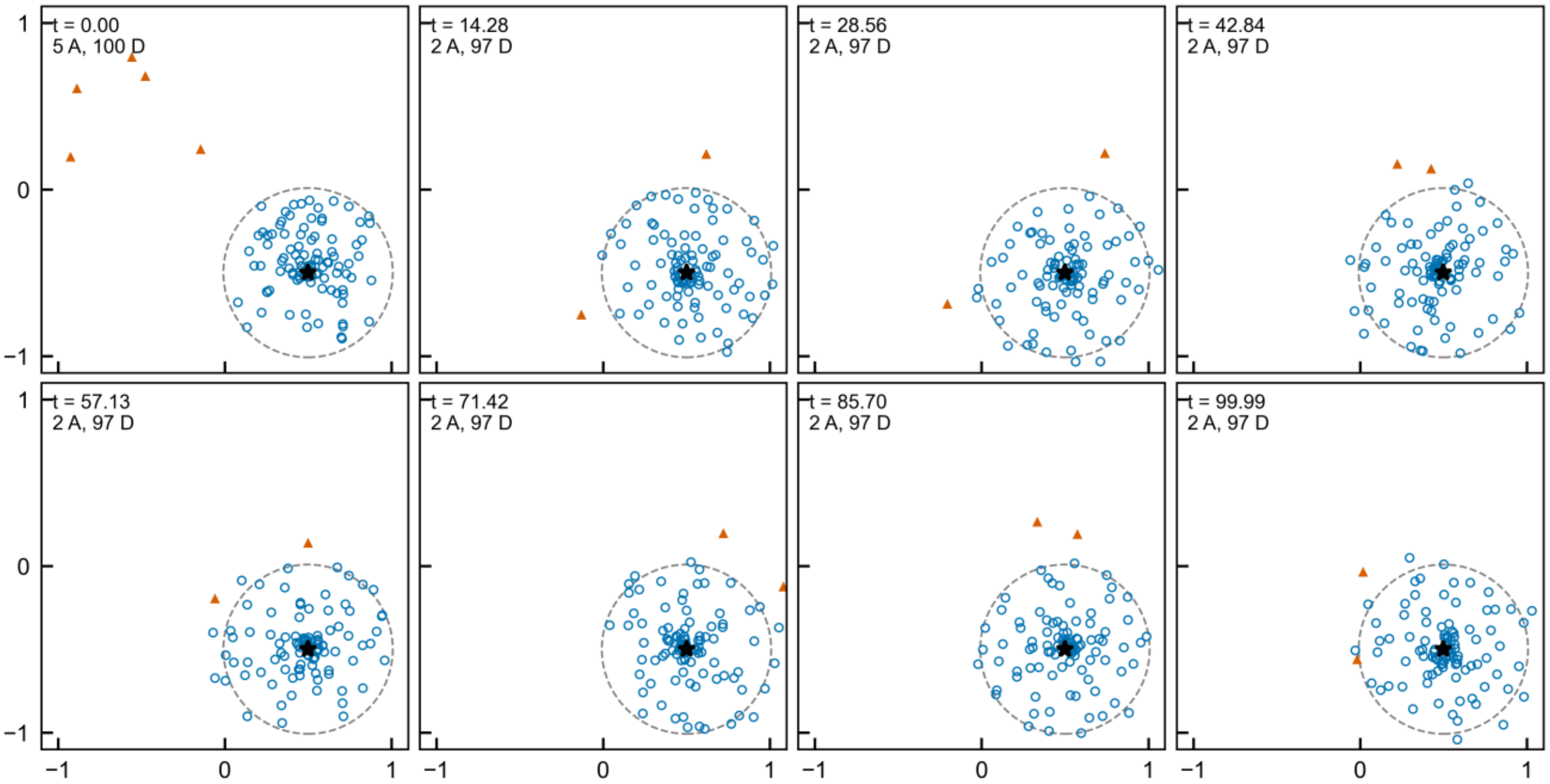

**FIG. 3.** Snapshots of an engagement at $N_A = 5$, $N_D = 100$ (triangles, attackers; circles, defenders; star, goal; dashed circle, sphere of influence). Three attackers are removed before t = 14; the two survivors orbit outside the sphere until $T_{max}$.

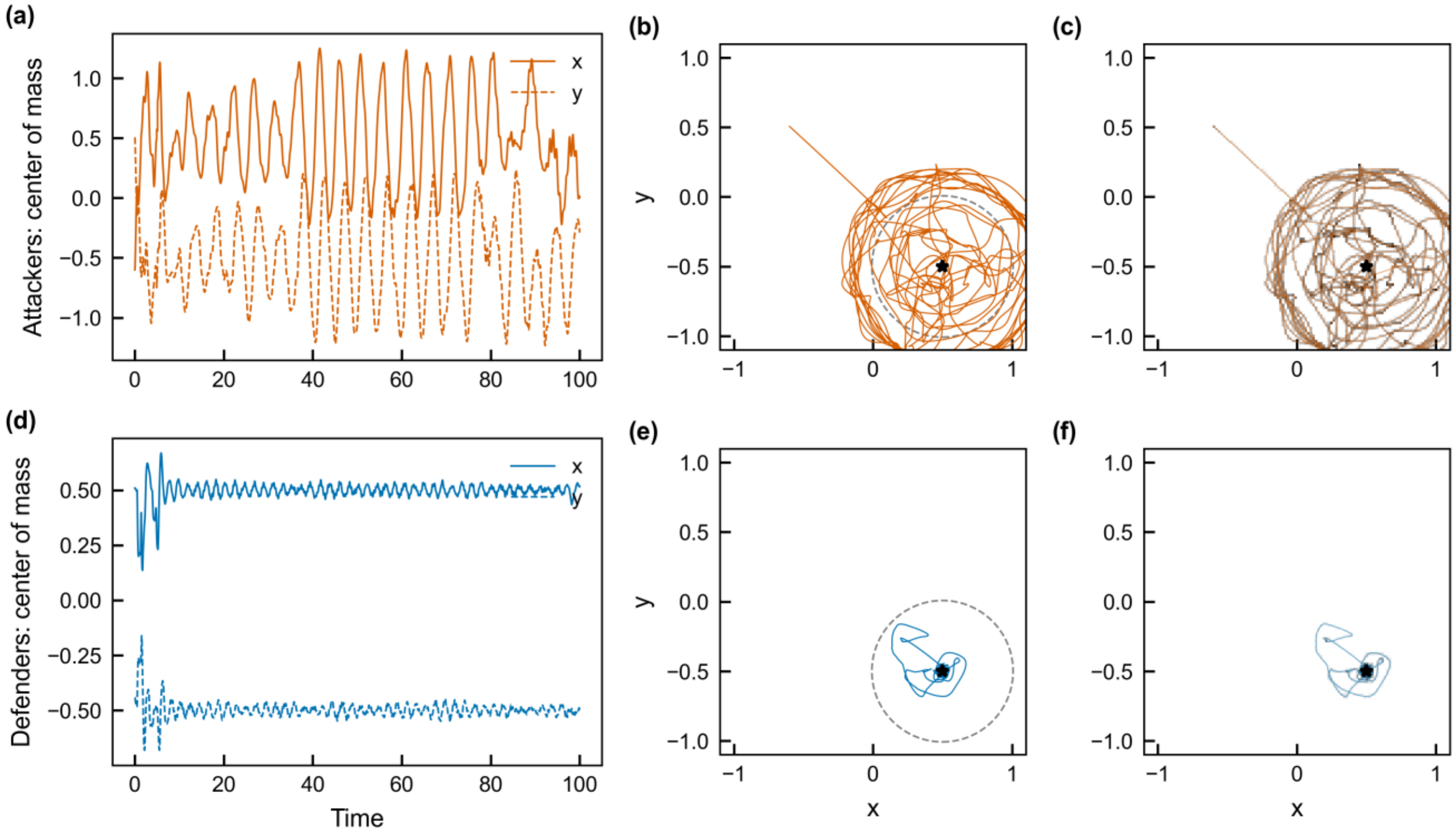

**FIG. 4.** The same engagement: center-of-mass time series (a, d), trajectories in the plane (b, e) and residence density of the center of mass (c, f) for the attackers (top) and the defenders (bottom).

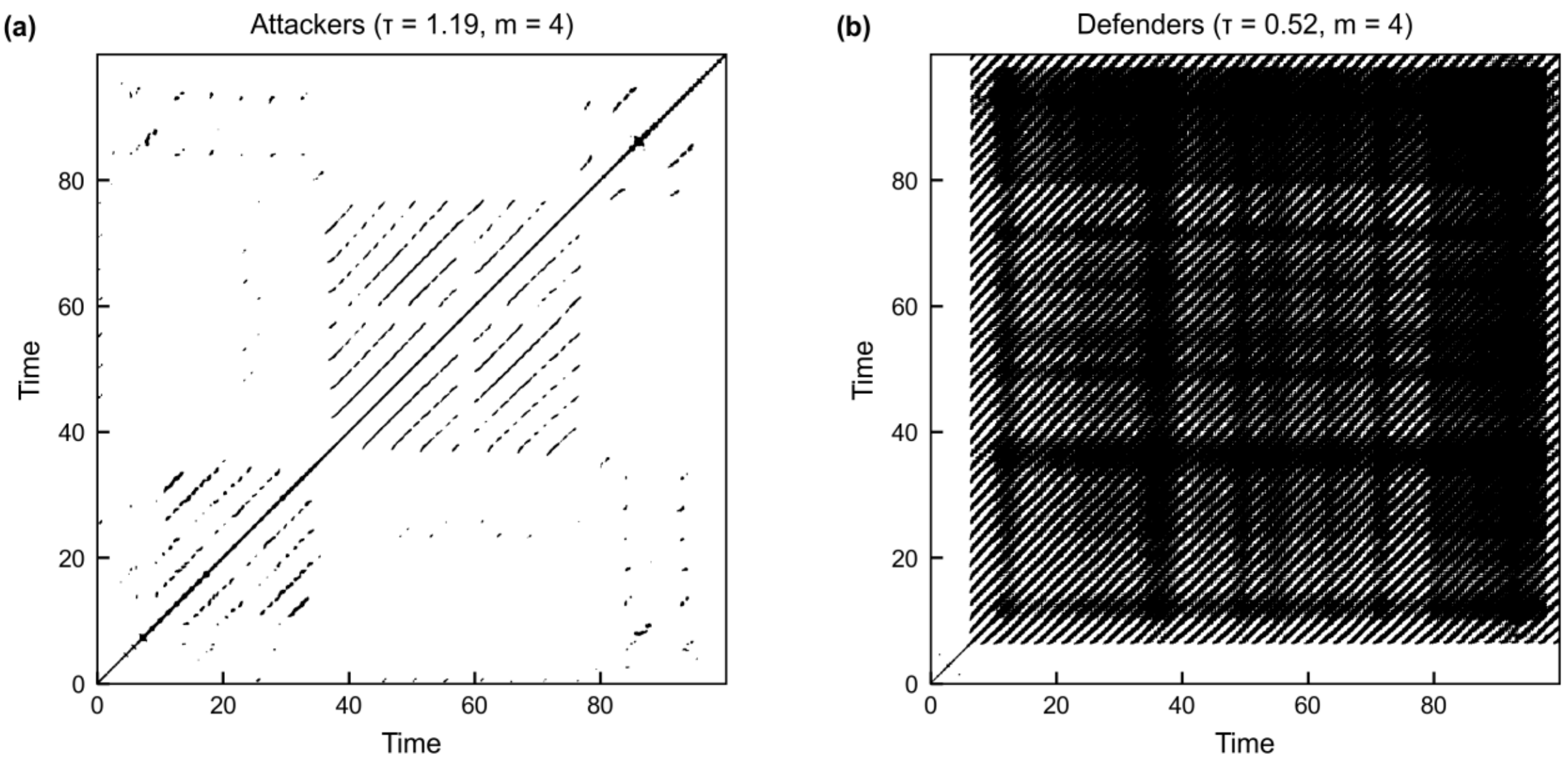

**FIG. 5.** Recurrence plots of the attacker (a) and defender (b) center-of-mass series of the same engagement (threshold 10% of the phase-space diameter; delay τ and dimension m as marked).

At parity (Fig. 6) the swarms engage from the outset and agents are removed in nearly equal numbers, consistent with one-to-one encounters between nearest adversaries (50 against 50 at the start, 38 against 38 at t = 2.2, 11 against 11 at t = 6.7); events that remove more than two agents also occur in the ensemble. The two center-of-mass series (Fig. 7(a)) move together with irregular, non-smooth fluctuations produced by the rapid succession of removals until, at t ≈ 11, one agent of each swarm is left; the pair then circle the goal on a periodic orbit, visible as the regular oscillation at the end of the series (the two centers nearly coincide there), until they meet and the defender prevails at t = 15.7. The residence density of both centers (Fig. 7(b)) is concentrated inside the sphere of influence. Far above parity ($N_A/N_D$ = 20, Fig. 7(c, d)) the attackers approach almost linearly; the attacker center of mass undergoes a slow, smooth oscillation about the goal, because the swarm is barely affected by the loss of a few members, while the defender center moves erratically as the five defenders engage attackers arriving from all sides, until the last defender is removed at t = 5.3.

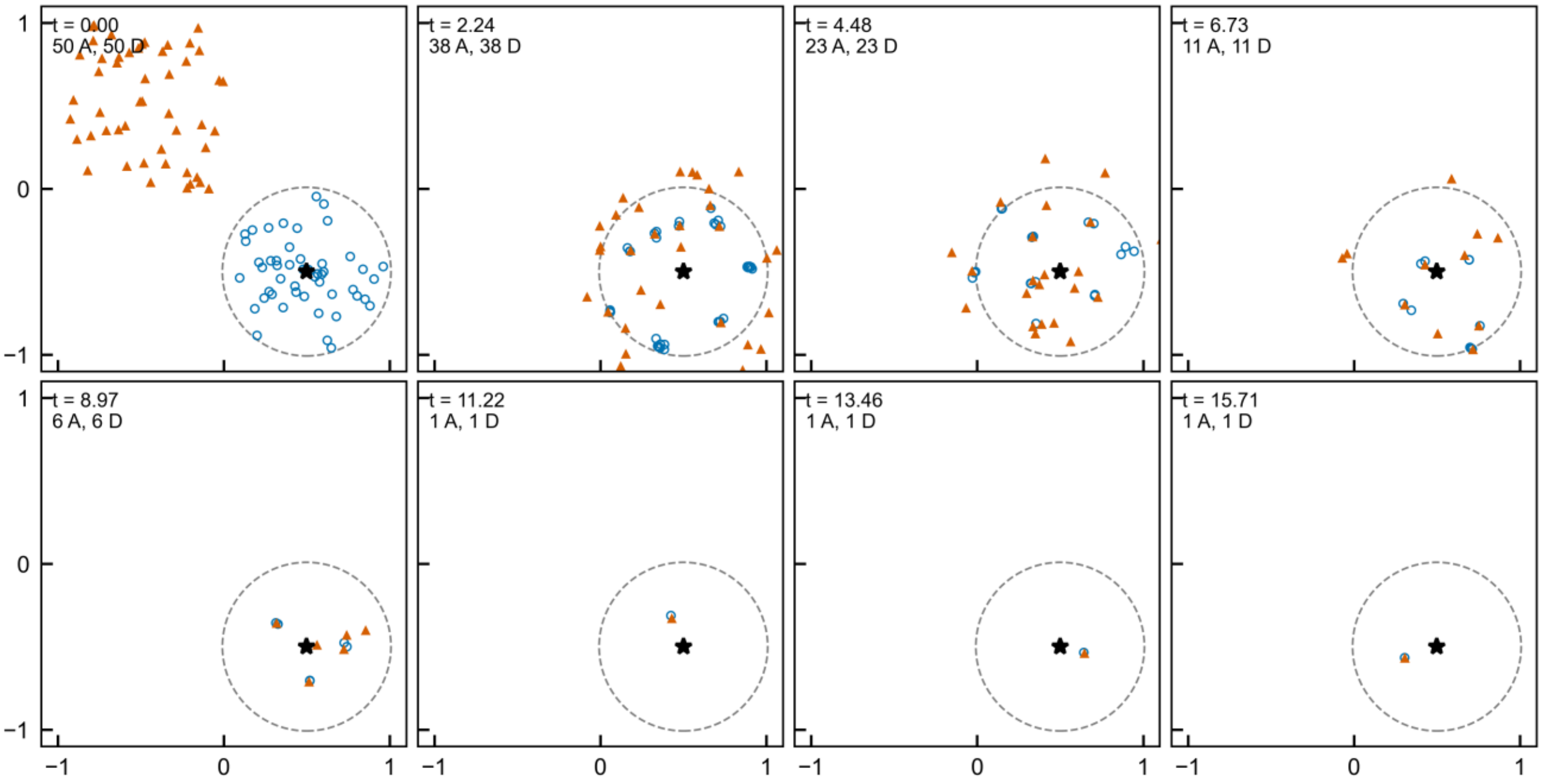

**FIG. 6.** Snapshots of an engagement at $N_A = N_D = 50$ at eight equally spaced times; the numbers of active agents are given in each panel.

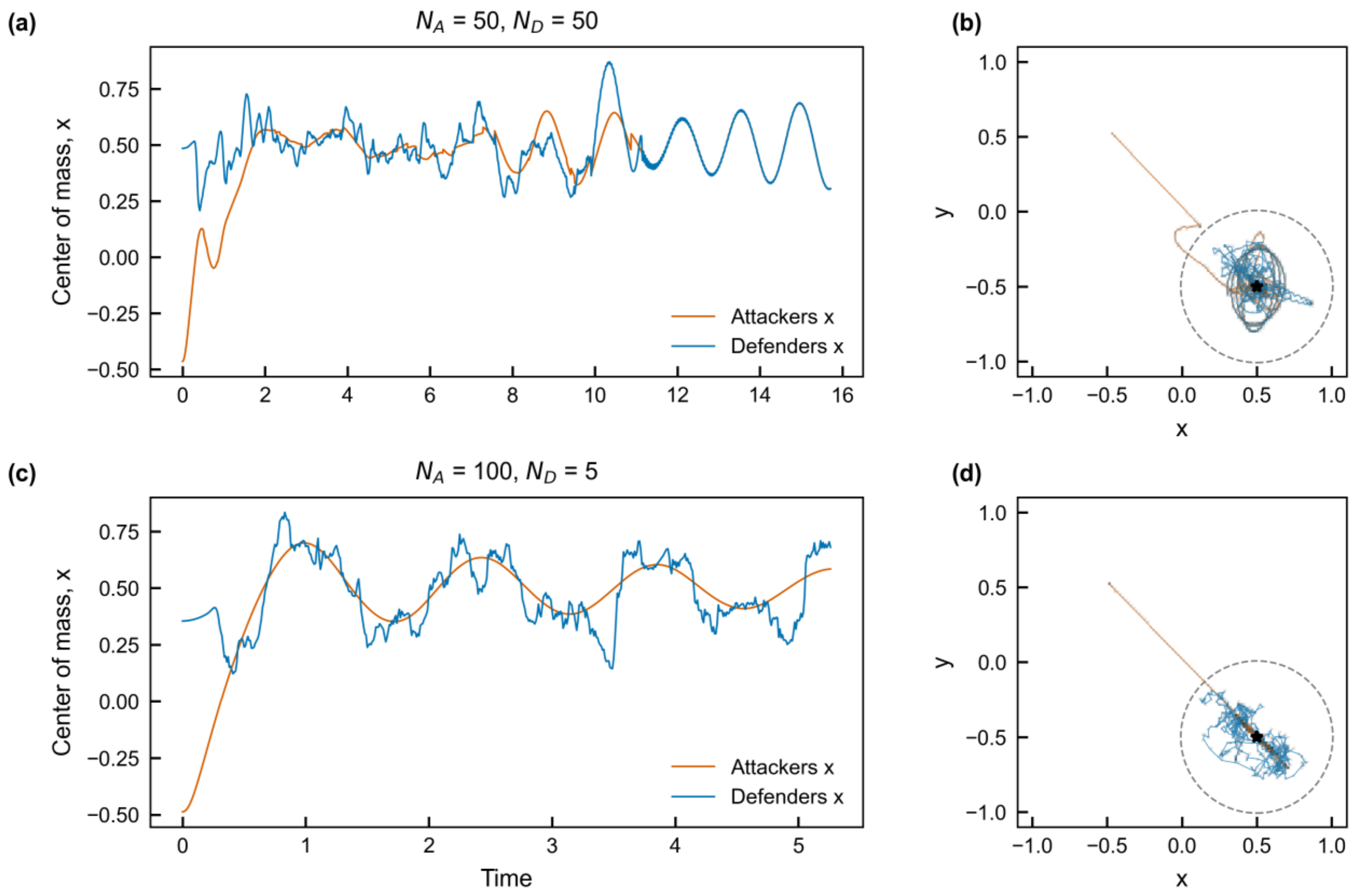

**FIG. 7.** Center-of-mass x component (a, c) and residence density of both centers of mass in the plane (b, d) for engagements at $N_A = N_D = 50$ (top) and $N_A = 100$, $N_D = 5$ (bottom).

### 4.3 Chaos: ensemble exponents, twin trajectories and the 0-1 test

Figure 8(a) shows the mean, minimum and maximum of $\lambda_1$ over the 1000 trials of each case for both swarms. The defender exponent has a positive mean at every ratio, between 0.005 and 0.037, with a maximum over trials of up to 0.57; the attacker exponent has a mean within ±0.013 of zero and is negative on average for every ratio above 0.75. Figure 8(b) resolves the distributions for the four cases whose per-trial values were retained. At $N_A/N_D = 0.05$ the two swarms are similar, with $\lambda_1 > 0$ in 55% and 53% of trials; the distributions are concentrated at zero with a positive tail, consistent with the transient chaos described in Section 4.2. At $N_A/N_D = 0.31$ and 1.0 the swarms separate: $\lambda_1 > 0$ in 97% and 98% of trials for the defenders, with medians of 0.025 and 0.013, against 60% and 3% for the attackers, whose median is at or below zero. At $N_A/N_D = 20$ the defender exponent is positive in every trial. The largest exponent in the whole study, $\lambda_1 = 0.54$, occurs for the defenders at $N_A/N_D = 1.33$, where the recurrence plot of that trial shows the broken diagonals of chaotic motion.

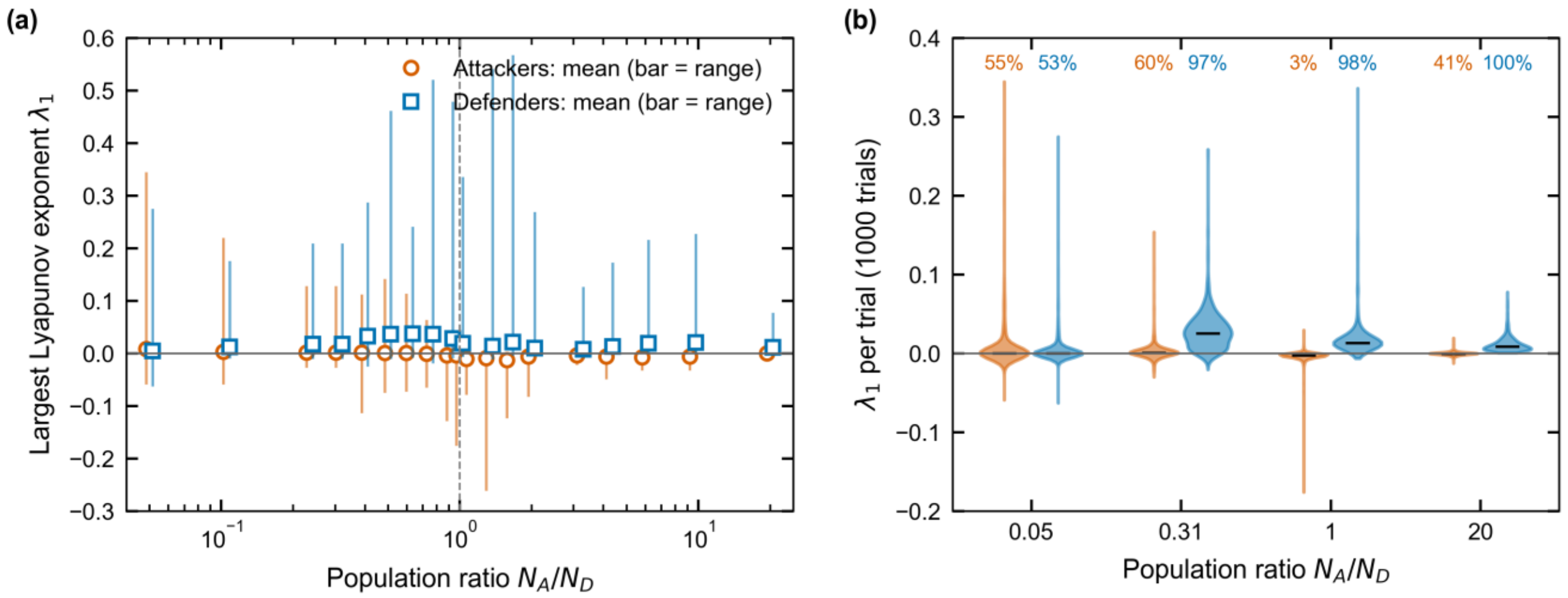

**FIG. 8.** (a) Mean (symbols) and range (bars) of the largest Lyapunov exponent $\lambda_1$ over 1000 trials for each case. (b) Distributions of $\lambda_1$ per trial for $N_A/N_D$ = 0.05, 0.31, 1 and 20; the percentage of trials with $\lambda_1 > 0$ is given above each distribution.

Figure 9 shows what the interacting dynamics do when they are not cut short by attrition. In the non-attritional variant the 0-1 test statistic lies between 0.97 and 1.00 for both swarms at every ratio tested (0.05, 0.5, 1 and 2, two initial conditions each): when the swarms interact continuously, the motion of both centers of mass is chaotic, including that of the five attackers chased by one hundred defenders at $N_A/N_D = 0.05$. The recurrence plots at parity (Fig. 9(e, f)) show only short diagonal segments and isolated points for both swarms. The swarms differ in the strength of the chaos rather than in its presence. For $N_A/N_D \geq 0.5$ the Rosenstein exponent of the defender series is three to five times that of the attackers (16 to 22 against 4 to 7 per unit time; both are upper estimates, because the nearest-neighbor divergence of a series that needs five to seven embedding dimensions saturates quickly), and the sample entropy of the defenders exceeds that of the attackers at every scale, with means over scales 1 to 20 of 1.80 to 1.92 against 1.18 to 1.31. The defender curve rises steeply and peaks at scales of 6 to 9, that is, at 0.12 to 0.18 time units, whereas the attacker curve rises monotonically over the whole range (Fig. 9(c)). At $N_A/N_D = 0.05$ the ordering reverses: the five attackers, chased without respite, carry the higher entropy (0.89 to 1.04 against 0.43 to 0.60) at comparable exponents.

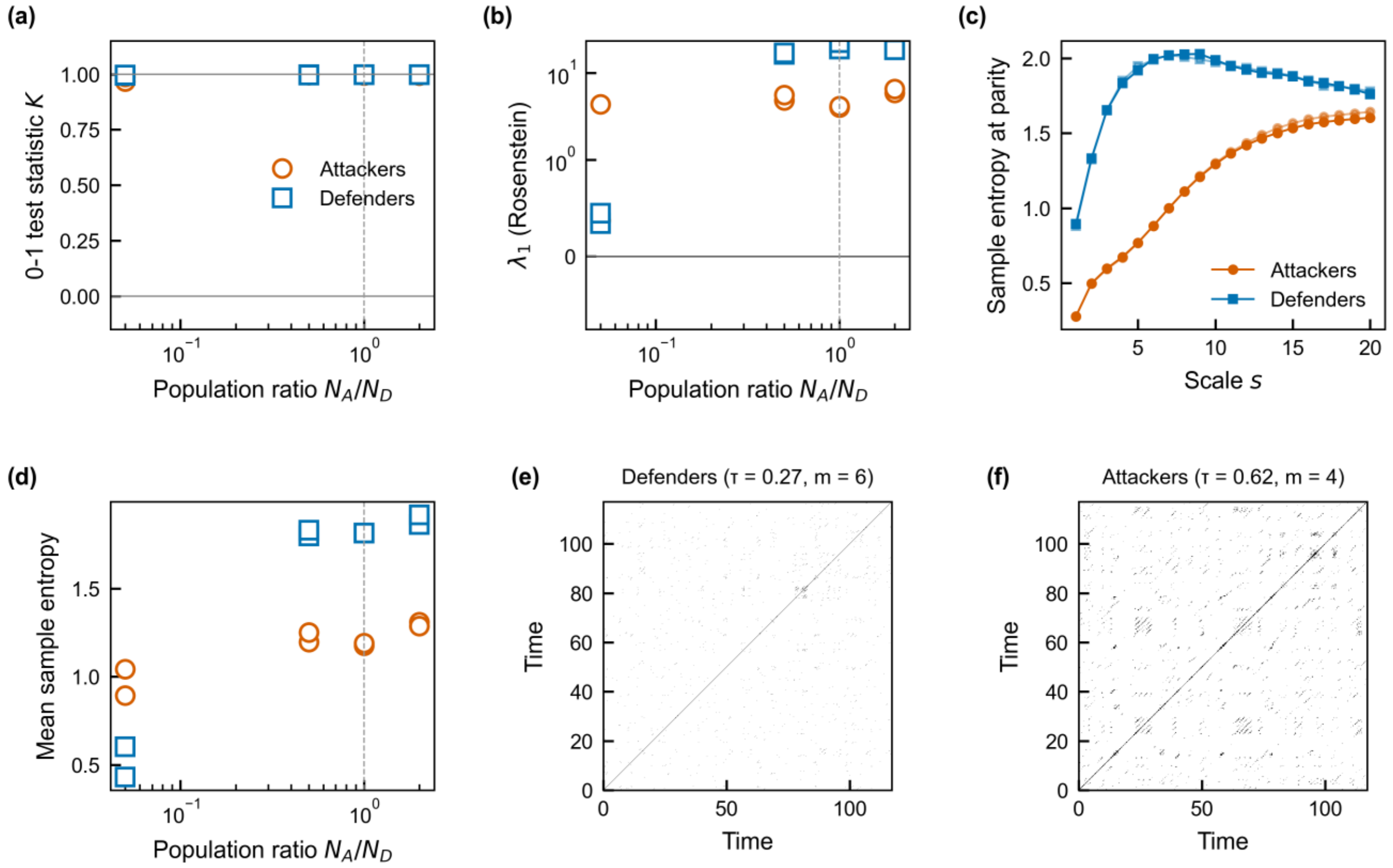

**FIG. 9.** Non-attritional engagements (agent removal switched off, 150 time units, first 30 discarded): (a) 0-1 test statistic K, (b) largest Lyapunov exponent $\lambda_1$ (Rosenstein) and (d) sample entropy averaged over scales 1 to 20 of the center-of-mass series of each swarm versus $N_A/N_D$, two initial conditions per ratio; (c) sample entropy versus scale at $N_A = N_D = 50$; (e, f) recurrence plots of the defender and attacker centers of mass at $N_A = N_D = 50$.

The twin trajectories (Fig. 10) show how this chaos acts in the attritional engagements. Before the first removal of an agent, which occurs a median 0.44 to 0.46 time units after the start at every ratio, the separation of the attacker centers of mass decays, at a median rate of −0.8 to −1.4 per unit time, while that of the defender centers grows, at 3.4 to 5.2 per unit time: the cohesive attacker swarm damps a displacement of one of its members on the way to the goal, whereas the defender swarm, orbiting in its Morse well with the attraction switching at the sphere of influence, amplifies it. After the first contact both separations grow by seven orders of magnitude within about one time unit, in steps that coincide with removals, and reach the swarm scale, $d = 10^{-3}$, at a median $t_h$ of 1.40 time units (interquartile range 1.26 to 1.54) at $N_A/N_D = 0.5$, 1.13 (1.02 to 1.21) at 0.91, 1.02 (0.96 to 1.14) at parity, 1.08 (0.99 to 1.17) at 1.1 and 0.90 (0.79 to 0.97) at 2; the defender horizons are the same to within 0.07. The horizon is thus about one time unit at every ratio, and shortest where the attackers' surplus is largest. Engagements last 7 to 25 time units, so their state is unpredictable beyond the first tenth of their duration. The $10^{-8}$ displacement changed the dominant swarm in 2 of 47 twin pairs at $N_A/N_D = 0.91$, 1 of 47 at parity and 3 of 45 at 1.1, that is in 6 of 139 pairs (4.3%) in the contested band, and in none of 50 pairs at 0.5 or 42 at 2, where the outcome is not in doubt.

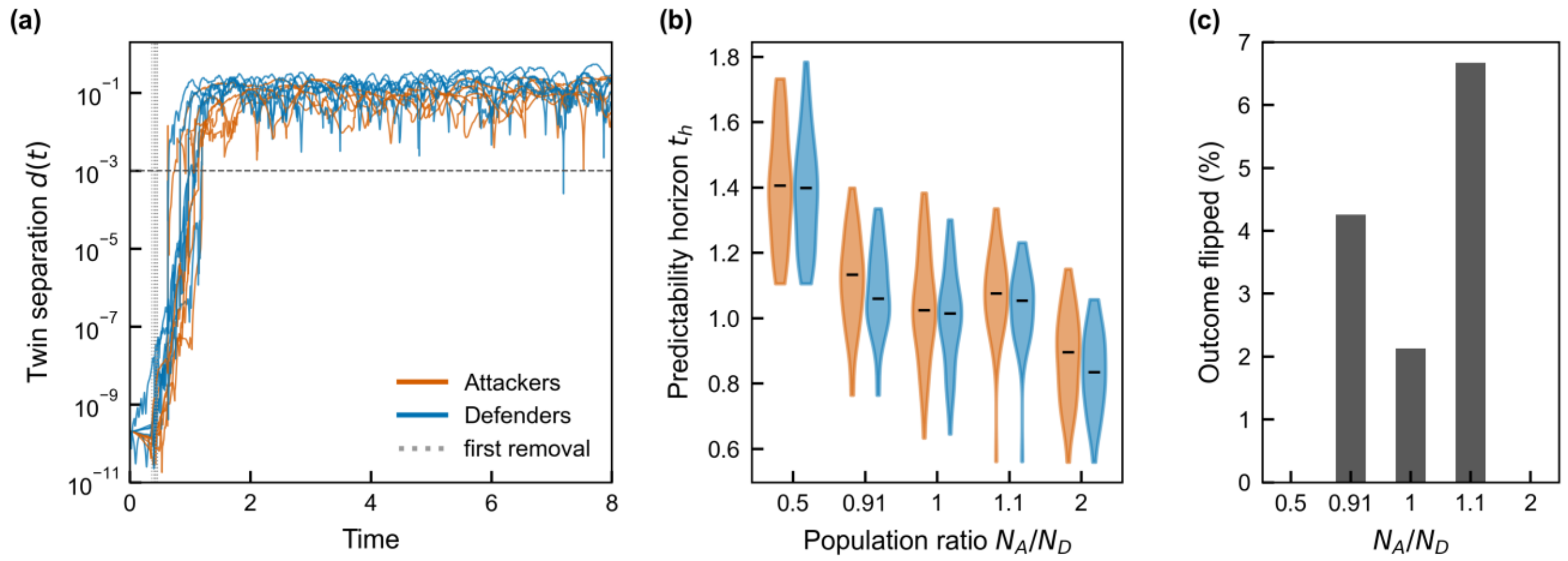

**FIG. 10.** Twin trajectories differing by a $10^{-8}$ displacement of one attacker and one defender: (a) separation of the twin centers of mass d(t) for six trials at $N_A = N_D = 50$ (dotted lines, first removal; dashed line, the horizon threshold $10^{-3}$); (b) distributions of the predictability horizon $t_h$ over 42 to 50 valid twin pairs per ratio; (c) percentage of twin pairs that ended with different dominant swarms.

The ensemble exponents and the sustained-interaction results are reconciled by the attrition. In an engagement at or above parity the attackers remove the defenders within 10 to 25 time units, and the surviving attackers relax to a periodic or quasi-periodic orbit around the goal; that orbit dominates the attacker center-of-mass series and gives a non-positive Wolf exponent, whereas the defender series, confined to the sphere of influence and reacting to attackers until the last removal, remains irregular (Figs. 4 and 7). The asymmetry of Fig. 8 is therefore a property of the attritional engagement and its resolution, not of the interacting dynamics as such; while the swarms interact, both are chaotic, and the goal-bound defenders are the more strongly chaotic component.

### 4.4 Complexity: multiscale entropy

Figure 11 shows the ratio of the multivariate multiscale entropy at scale 20 to that at scale 1 for both swarms, with error bars giving the standard deviation over trials. The ratio of the defenders exceeds that of the attackers at every population ratio, and both swarms have entropy increasing with scale, the signature of structure across temporal scales rather than of uncorrelated noise.[62] The attackers' ratio is largest at the smallest population ratio (1.14 at $N_A/N_D = 0.05$, where the few surviving attackers orbit the sphere of influence for the whole run), falls to 1.04 across the contested band below parity, rises to a secondary maximum of 1.09 at parity, and settles at 1.01 for $N_A/N_D \geq 1.6$, where the attackers overwhelm the defenders quickly and their center of mass barely moves. The defenders' ratio is 1.33 at $N_A/N_D = 0.05$, lies between 1.12 and 1.23 for $0.07 \leq N_A/N_D \leq 0.91$, rises to 1.36 at parity and stays between 1.35 and 1.42 up to $N_A/N_D = 4.25$, before falling to 1.25 at $N_A/N_D = 6$ and 1.11 at 20; it is most variable from trial to trial for $1 \leq N_A/N_D \leq 1.33$, the cases that straddle the outcome transition, where the engagement lasts longest, the exponents are most widely distributed and both outcomes occur. The ordering of the swarms by entropy matches their ordering by Lyapunov exponent: the defenders, constrained to a small perimeter and forced to engage attackers arriving from several directions, are the less ordered swarm, and their complexity is highest where the forces on them are most nearly balanced.

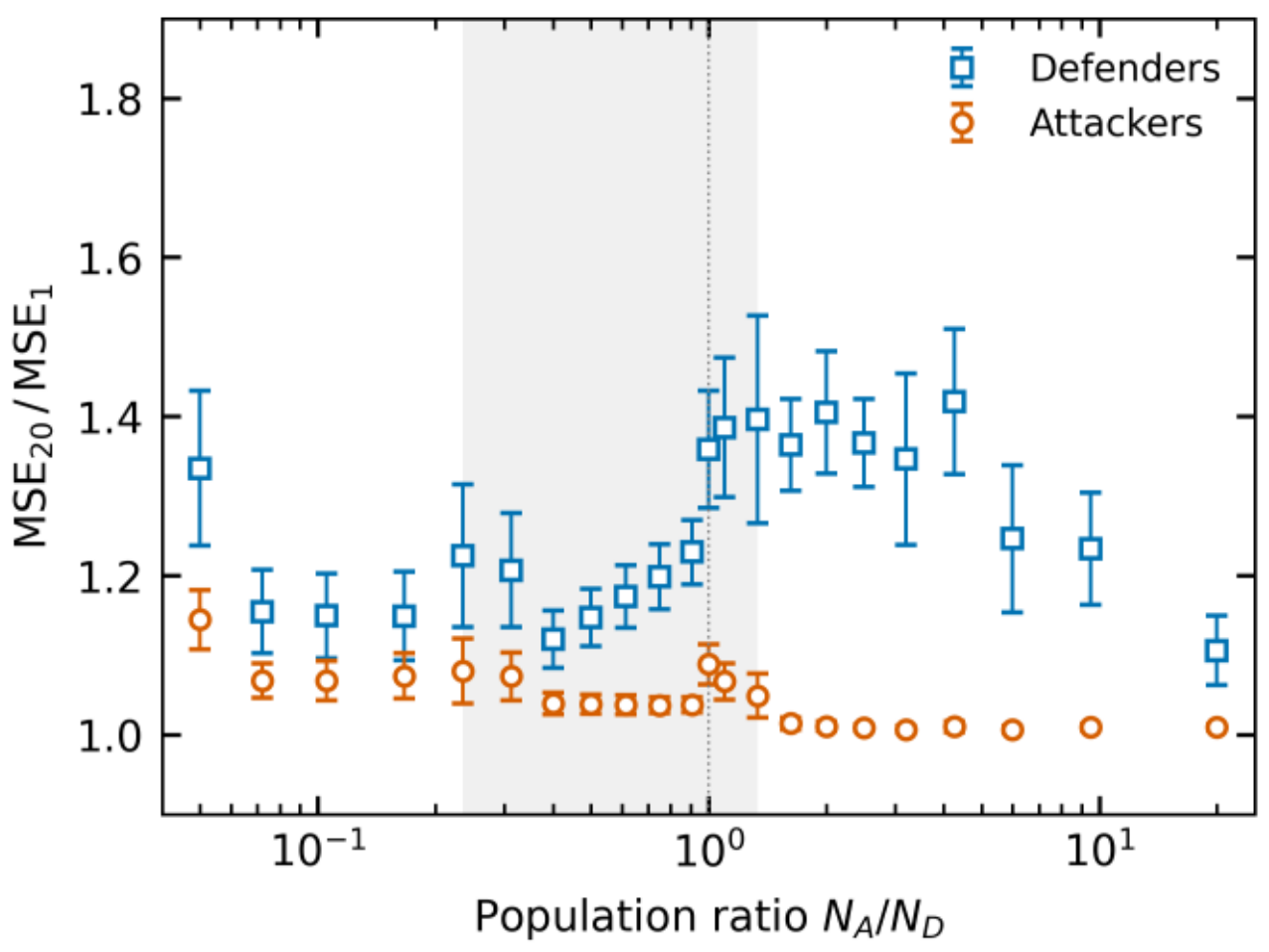

**FIG. 11.** Ratio of the multivariate multiscale entropy at scale 20 to that at scale 1 for the two-component center-of-mass series of each swarm, for all 22 cases; error bars are the standard deviation over trials (1000 per case, except 600 at $N_A/N_D = 1.63$ and 905 at 2). The shaded band is the contested band of Fig. 2.

### 4.5 Sensitivity to a finite sensing radius

Figure 12 compares the original model with the two finite-sensing-radius variants. Far from parity the sensing radius makes no difference to the outcome: at $N_A/N_D = 0.5$ the defenders annihilate the attackers in every trial with $R_s = \infty$ and with $R_s = 0.5$, in a median 5.3 instead of 6.9 time units, and at $N_A/N_D = 2$ the attackers dominate in 49 of 49 and 47 of 49 trials respectively, with the same median time (18 and 17 time units); the two exceptions at $R_s = 0.5$ are stand-offs in which two surviving defenders held the goal. Near parity the picture changes in two steps. With $R_s = 0.5$ the attackers' share at $N_A/N_D = 1.1$ falls from 96% to 62% (Wilson 95% interval 48 to 74%), not because the defenders win more engagements by attrition, which they do in none, but because 38% of the trials end as stand-offs: the attrition stops with a few survivors on each side that no longer sense one another, and the goal is held to $T_{max}$. At parity the same mechanism turns 44% of the trials into stand-offs, against none with $R_s = \infty$, so that the median time to outcome is almost unchanged (16 against 14 time units) while the mean rises from 14 to 52; at $N_A/N_D = 0.91$ the defenders still annihilate the attackers in 92% of trials. With $R_s = 0.25$ the engagement changes character. At parity and above, attrition no longer resolves it: no trial ends by annihilation of either swarm, 90% and 72% of trials end as stand-offs at $N_A/N_D = 1.0$ and 1.1, and the remaining 10% and 28% are breaches. Where the defenders hold the surplus ($N_A/N_D = 0.91$) they still annihilate the attackers in 64% of trials, but 24% end as stand-offs and 12% as breaches, at a ratio where long-range sensing gives the attackers no chance at all. The breaches are fast (mean 4 to 8 time units): with short-range sensing an attacker that approaches the goal from a direction not covered by a defender is neither repelled on the way nor pursued until it is within 0.25 of a defender, and it reaches the goal before the defenders react. The multiscale entropy of the two swarms also changes order at small $R_s$ (Fig. 12 does not show this): in the stalled engagements the attacker center of mass becomes nearly stationary and its entropy ratio collapses, while the defenders keep orbiting the goal.

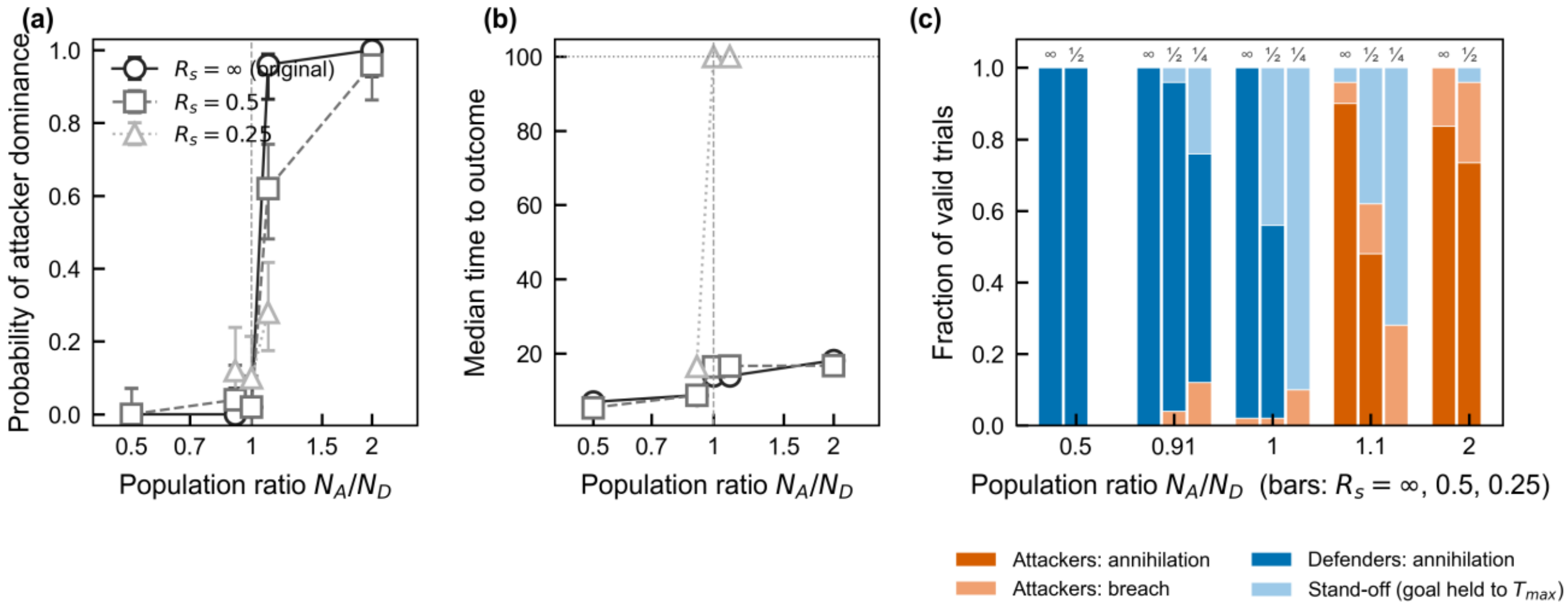

**FIG. 12.** Effect of a finite sensing radius $R_s$ on the attritional engagement, 49 or 50 valid trials per condition: (a) probability of attacker dominance with Wilson 95% intervals and (b) median time to outcome for $R_s = \infty$ (original model), 0.5 and 0.25; (c) composition of the outcomes (attacker dominance by annihilation of the defenders or by breach of the goal; defender dominance by annihilation of the attackers or by a stand-off in which the goal is held to $T_{max}$), one bar per sensing radius at each ratio.

The sharp outcome transition and the attritional character of the engagement are therefore properties of long-range sensing. The exponential decay of the Morse force confines most of the intra-swarm interaction to a few length scales, so a sensing radius of 0.25 removes little of it; the inter-swarm forces decay only as the inverse distance, and the sensing radius removes a large share of them. What is removed is precisely the mechanism of the parity transition, the mutual engagement of every attacker with every defender that converts a numerical surplus into annihilation. With short-range sensing the outcome depends on the geometry of interception rather than on numbers, which is the regime addressed by the interception and herding strategies of the control literature.[25,27]

## 5. DISCUSSION

The central result is the sharpness of the outcome transition. A change of five agents in a swarm of fifty takes the attackers from a 5% to a 97% chance of success. The mechanism is visible in the time-to-outcome data and in the snapshots. Below parity, every attacker that enters the sphere of influence draws a defender, and the defenders' numerical surplus is spent removing attackers while the goal is screened by the defenders that remain; the surplus is recovered after every one-to-one removal, so the defenders keep their advantage to the end. Above parity the attackers keep the surplus, and once the defenders are reduced below the number needed to screen the goal, the remaining attackers either breach it or finish the defenders. Because removals are predominantly one-to-one, the sign of $N_A - N_D$ is nearly conserved through the engagement, which is why the ratio, not the absolute numbers, decides the outcome and why the transition is concentrated at parity. The transition resembles the critical transition of colliding swarms[39] in that a small change of a control parameter switches the collective state, but the control parameter here is a population ratio and the switch is between outcomes of an attritional engagement rather than between milling and scattering. It also recalls the prey-survival transitions found as functions of interaction range[36] and predator-to-prey mass ratio,[37] in which a narrow intermediate regime separates certain capture from certain survival; in our model that regime is the contested band.

The second result concerns chaos and predictability, and three independent measurements give a consistent picture. While the swarms interact, the motion of both is chaotic: the 0-1 test gives $K \approx 1$ for both centers of mass at every ratio in sustained interaction. The goal-bound defenders are the more strongly chaotic component for $N_A/N_D \geq 0.5$, by divergence rate and by entropy, and the twin trajectories show why: before contact a small displacement decays in the cohesive attacker swarm, which is driven toward the goal by a smooth linear force and held together by its Morse interactions, but grows in the defender swarm, which orbits in a Morse well around the goal and whose attraction to the attackers switches on when an attacker crosses the sphere of influence and off when it leaves. The defender subsystem is a piecewise-smooth, event-driven system, in which switching and attrition continually re-set the local equilibria of the center of mass, as seen in the multiple attractors of Figs. 4 and 7. Once contact is made, removals drive the separation of the twin realizations to the swarm scale within about one time unit, whatever the ratio: the attritional engagement has a predictability horizon of order one time unit against a duration of 7 to 25, and near parity a perturbation of one part in $10^8$ changes the winner in a few percent of trials. The ensemble of Wolf exponents, which found the defenders chaotic in nearly every trial and the attackers in few, records the resolution of the engagement rather than the interaction: the surviving attackers relax to a regular orbit, the constrained defenders do not. Kamimoto et al.[40] found that the milling state formed by two colliding swarms becomes chaotic as the inter-swarm repulsion is varied, and Medeiros and Feudel[42] found chaos during the transitions between collective states of a single Morse–Rayleigh swarm; the present model shows that in an asymmetric engagement the strength of the chaos differs between the two swarms, and identifies the goal constraint and the switching pursuit as the cause.

The ensemble exponents are small, of order $10^{-2}$ in units of inverse simulation time, because they are averaged over series that end in regular orbits, and the original version of this study described the system as being at the edge of chaos on that basis.[68-70] The sustained-interaction exponents and the twin-trajectory horizons show that the interacting dynamics themselves are strongly chaotic, with e-folding times of order 0.05 to 0.3 time units. The more specific statement we now make is that the engagement is chaotic while the swarms interact and regular once it resolves, and that the transient chaos seen at the smallest ratios, where both swarms have positive ensemble exponents in about half the trials, is the same phenomenon seen in engagements that last long enough for the exponent to register it.

The finite-sensing-radius study answers the objection that Morse-type models couple every agent to every other, and the answer is not that the coupling does not matter. Far from parity it does not; near parity it decides the character of the engagement. With long-range sensing every attacker engages every defender, a numerical surplus is converted into annihilation, and the outcome is decided by numbers within a 10% band around parity. With $R_s = 0.25$ the swarms meet only locally; at and above parity attrition stops with survivors on both sides and the outcome is either a fast breach by an attacker that was never repelled or pursued, or a stand-off, while below parity the defenders' surplus still annihilates the attackers in two trials out of three. Section 4.5 gives the numbers.

For the design of defensive swarms the results therefore have a two-part reading. When sensing and communication are long-range, numbers decide: a defending swarm that cannot maintain at least parity after the first exchanges will lose the goal in nearly every engagement, and one that can will keep it. When sensing is short-range, numbers no longer decide, and the outcome turns on the geometry of interception, which is the regime addressed by the interception and herding strategies of Chipade and Panagou[27] and the optimal-control defenses of Walton et al..[25] The same two regimes are recognizable in the biological case, in which group size determines the success of group hunting[24] but only when the hunters can see the prey.

The model has limitations that define the scope of the conclusions. It is two-dimensional, deterministic apart from the initial conditions, and without communication delay or sensing noise, both of which are known to produce bifurcations of swarm patterns.[71,72] The goal is fixed, the agents are identical within a swarm, and the inter-swarm constants were chosen for a balanced engagement rather than fitted to any biological or engineered system. The Monte Carlo ensemble covers initial conditions but not parameter variation. Extensions to three dimensions, to noisy and delayed sensing, to heterogeneous agents, and to a moving or multiple goals are natural next steps, and the sharpness of the parity transition is the first property we would test under them.

# 6. CONCLUSION

We have studied an attacker–defender swarm model as a dynamical system, with Morse and inverse-distance potentials, Rayleigh self-propulsion, a goal-bound defender swarm and attrition on contact. From 22 population ratios and 22,000 trials we conclude:

1. The outcome is decided by the population ratio in a narrow band around parity: the probability of attacker dominance rises from 0.046 at $N_A/N_D = 1$ to 0.971 at 1.1, and both outcomes are possible only for $0.24 \leq N_A/N_D \leq 1.33$. Engagements last longest just above parity.
2. The interacting dynamics are chaotic for both swarms (0-1 test $K \geq 0.97$ at every ratio in sustained interaction), and the goal-bound defenders are the more strongly chaotic component for $N_A/N_D \geq 0.5$. Attritional engagements become unpredictable within about one time unit of their start, a $10^{-8}$ displacement of one agent flips the outcome in a few percent of trials near parity, and before contact the attackers damp such a perturbation while the defenders amplify it. The asymmetry of the ensemble Wolf exponents (defenders positive in 97 to 100% of trials, attackers in few at and above parity) records the attackers' relaxation to a regular orbit once the engagement resolves.
3. The defenders are the more complex swarm by multiscale entropy at every population ratio in the original ensemble and for $N_A/N_D \geq 0.5$ in sustained interaction; their complexity is highest between parity and $N_A/N_D \approx 4$, and the attackers' shows a secondary maximum at parity.
4. The sharp transition is a property of long-range sensing. A sensing radius of 0.5 leaves the outcomes far from parity unchanged but turns about 40% of the engagements at and just above parity into stand-offs; at 0.25, attrition no longer resolves the engagement at or above parity, and the outcome there is a quick breach of the goal in 10 to 28% of trials and a stand-off otherwise.

Together these results show that the physics of interacting self-propelled swarms, developed for symmetric collisions, extends to asymmetric, attritional engagements, and that in such engagements the collective dynamics of the two sides can differ qualitatively even when the agents are identical.

# ACKNOWLEDGMENTS

The authors thank the staff of the University of Alabama High-Performance Computing group for their support.

# AUTHOR DECLARATIONS

Conflict of interest: The authors have no conflicts to disclose.

Author contributions: Soham Gupta: Conceptualization (equal); Methodology (lead); Software (lead); Investigation (lead); Formal analysis (lead); Visualization (lead); Writing – original draft (lead); Writing – review and editing (equal). John Baker: Conceptualization (equal); Methodology (supporting); Supervision (lead); Resources (lead); Writing – review and editing (equal).

# DATA AVAILABILITY

The simulation code, the configuration files, the outcome and Lyapunov-exponent data for all 22 cases, the per-trial data of the sensitivity study, and the analysis scripts that produce every figure are available from the corresponding author upon reasonable request.

# APPENDIX: CARTESIAN COMPONENTS OF THE FORCES

This appendix derives the Cartesian force components that enter Eqs. (1) to (4). The force on each agent is the negative gradient of the potentials of Section 2.1. Following Kolon and Schwartz,[38] the interaction between agents of the same swarm is described by the scaled Morse potential

$$\phi_{ij} = C\exp(-\frac{|r_{ij}|}{l}) - \exp(-|r_{ij}|) \tag{17}$$

where $C$ defines the depth of the repulsive potential well and $l$ is a constant that sets the ratio of the repulsive to attractive length scales. Since the force is the negative gradient of the potential,

$$\vec{F} = -\vec{\nabla}\phi_{ij} \tag{18}$$

Because the potential is radially symmetric, its Cartesian force components follow from the chain rule,

$$\begin{aligned} F_{x,ij} &= -\frac{\partial\phi_{ij}}{\partial x_{ij}} = -\frac{\partial\phi_{ij}}{\partial r_{ij}}\frac{\partial r_{ij}}{\partial x_{ij}} \\ &= [-\frac{C}{l}C\exp(-\frac{|r_{ij}|}{l}) - \exp(-|r_{ij}|)]\frac{\partial r_{ij}}{\partial x_{ij}} \end{aligned} \tag{19}$$

$$\begin{aligned} F_{y,ij} &= -\frac{\partial\phi_{ij}}{\partial x_{ij}} = -\frac{\partial\phi_{ij}}{\partial r_{ij}}\frac{\partial r_{ij}}{\partial y_{ij}} \\ &= [-\frac{C}{l}C\exp(-\frac{|r_{ij}|}{l}) - \exp(-|r_{ij}|)]\frac{\partial r_{ij}}{\partial y_{ij}} \end{aligned} \tag{20}$$

where the derivatives of the separation with respect to the coordinates are

$$r_{ij} = \sqrt{(x_i - x_j)^2 + (y_i - y_j)^2} \tag{21}$$

Differentiating the potential and substituting these derivatives gives

$$F_{x,ij,M} = [-\frac{C}{l}\exp(\frac{|r_{ij}|}{l}) + \exp(-|r_{ij}|)]\frac{(x_i - x_j)}{r_{ij}} \tag{22}$$

$$F_{y,ij,M} = [-\frac{C}{l}\exp(\frac{|r_{ij}|}{l}) + \exp(-|r_{ij}|)]\frac{(y_i - y_j)}{r_{ij}} \tag{23}$$

which are the Cartesian components of the Morse force that produces swarming in both the attackers and the defenders, each swarm with its own constants (Table 1).

The goal force on the attackers and the inter-swarm attraction and repulsion derive from the potentials of Espitia and Sofrony.[46] The goal potential is

$$\phi_G = -k_{obj} r_{iG}^2 \tag{24}$$

where the distance between the goal and the agent is $r_{iG} = \sqrt{(x_i - x_G)^2 + (y_i - y_G)^2}$ . The Cartesian components of the goal force follow from the gradient of the potential along each direction,

$$F_{x,G} = [2k_{obj} r_{iG}] \frac{\partial r_{iG}}{\partial x_{iG}} = (\frac{x_i - x_G}{r_{iG}}) \tag{25}$$

$$F_{y,G} = (\frac{y_i - y_G}{r_{iG}}) \tag{26}$$

The goal force is attractive and grows linearly with the distance to the goal. The inter-swarm potential of Espitia and Sofrony,[46] attractive or repulsive according to the sign of its constant, is

$$\vec{F}_{A/R} = -\vec{\nabla}(\pm k_{att/rep} r_{ij}^{-1}) \tag{27}$$

with Cartesian components

$$F_{x,A/G} = [\pm k_{obj} r_{ij}^{-2}](\frac{x_i - x_y}{r_{ij}}) \tag{28}$$

$$F_{y,A/G} = [\pm k_{obj} r_{ij}^{-2}](\frac{y_i - y_y}{r_{ij}}) \tag{29}$$

With the subscripts A and D for attackers and defenders, and $N_A$ and $N_D$ for their numbers, Newton's second law for each agent reads, in Cartesian components,

$$\frac{\partial^2 x_{A_i}}{\partial t^2} = \frac{1}{m_{A_i}} \sum_{\substack{A_i=1 \\ A_i \neq A_k}}^{N_A} [-\frac{C_A}{l_A} e^{-\frac{r_{A_i} r_{A_k}}{l}} + e^{-r_{A_i} r_{A_k}}] \frac{(x_{A_i} - x_{Ak})}{r_{A_i} r_{Ak}}$$
$$+ \frac{1}{m_{A_i}} \sum_{A_i=1}^{N_D} -\nabla[-k_{rep} r_{A_i D_j}^{-2}] \frac{(x_{A_i} - x_{D_j})}{r_{A_i D_j}} + \tag{30}$$
$$\frac{1}{m_{A_i}} \sum_{D_k=1}^{N_D} -\nabla[-k_{obj} r_{A_i G}] \frac{(x_{A_i} - x_G)}{r_{A_i G}} + (\alpha_{A_i} - \beta_{A_i} |\vec{v}_{A_i}|^2) \vec{v}_{A_i}$$

$$\frac{\partial^2 y_{A_i}}{\partial t^2} = \frac{1}{m_{A_i}} \sum_{\substack{A_i=1 \\ A_i \neq A_k}}^{N_A} [-\frac{C_A}{l_A} e^{-\frac{r_{A_i} r_{A_k}}{l}} + e^{-r_{A_i} r_{A_k}}] \frac{(y_{A_i} - y_{Ak})}{r_{A_i} r_{Ak}}$$
$$+ \frac{1}{m_{A_i}} \sum_{A_i=1}^{N_D} -\nabla[-k_{rep} r_{A_i D_j}^{-2}] \frac{(y_{A_i} - y_{D_j})}{r_{A_i D_j}} + \tag{31}$$
$$\frac{1}{m_{A_i}} \sum_{D_k=1}^{N_D} -\nabla[-k_{obj} r_{A_i G}] \frac{(y_{A_i} - y_G)}{r_{A_i G}} + (\alpha_{A_i} - \beta_{A_i} |\vec{v}_{A_i}|^2) \vec{v}_{A_i,x}$$

$$\frac{\partial x_{A_i}}{\partial t} = v_{A_i,x} \tag{32}$$

$$\frac{\partial y_{A_i}}{\partial t} = v_{A_i,y} \tag{33}$$

$$\frac{\partial^2 x_{D_i}}{\partial t^2} = \frac{1}{m_{D_i}} \sum_{\substack{D_i=1 \\ D_i \neq D_k}}^{N_D} [-\frac{C_D}{l_D} e^{-\frac{|r_{D_i} r_{D_k}|}{l}} + e^{-_{D_i} r_{D_k}}] \frac{(x_{A_i} - x_{A_k})}{r_{D_i} r_{D_k}}$$
$$+ \frac{1}{m_{D_i}} \sum_{D_i=1}^{N_A} -\nabla[-k_{obj} r_{D_i A_j}^{-2}] \frac{(x_{D_i} - x_{A_j})}{r_{D_i A_j}} + \tag{34}$$
$$[-\frac{C_{DG}}{l_{DG}} e^{-\frac{|r_{D_i} r_G|}{l_{DG}}} + e^{-|r_{D_i} r_G|}] \frac{(x_{D_i} - x_G)}{r_{D_i G}} + (\alpha_D - \beta_D |\vec{V}_{D_i}|^2) V_{D_i,x}$$

$$\frac{\partial^2 y_{D_i}}{\partial t^2} = \frac{1}{m_{D_i}} \sum_{\substack{D_i=1 \\ D_i \neq D_k}}^{N_D} [-\frac{C_D}{l_D} e^{-\frac{|r_{D_i} r_{D_k}|}{l}} + e^{-D_i r_{D_k}}] \frac{(y_{Ai} - y_{Ak})}{r_{D_i} r_{D_k}}$$

$$+ \frac{1}{m_{D_i}} \sum_{D_i=1}^{N_A} -\nabla[-k_{obj} r_{D_i A_j}^{-2}] \frac{(y_{D_i} - y_{A_j})}{r_{D_i A_j}} + \tag{35}$$

$$[-\frac{C_{DG}}{l_{DG}} e^{-\frac{|r_{D_i} r_G|}{l_{DG}}} + e^{-|r_{D_i} r_G|}] \frac{(y_{D_i} - y_G)}{r_{D_i G}} + (\alpha_D - \beta_D |\vec{V}_{D_i}|^2) V_{D_{y,i}}$$

$$\frac{\partial x_{D_i}}{\partial t} = v_{D_{i,x}} \tag{36}$$

$$\frac{\partial y_{D_i}}{\partial t} = v_{D_{i,y}} \tag{37}$$

Equations (30) to (37) are the scalar form of the governing Eqs. (1) to (4), subject to the initial conditions $(v_{A_{i,x}}, v_{A_{i,y}}), (x_{A_i}, y_{A_i}), (v_{D_{j,x}}, v_{D_{j,y}}), (x_{D_j}, y_{D_j})$, i.e., the initial positions and velocities of every agent in both swarms, and are integrated with the fourth-order Runge–Kutta scheme of Section 2.3.